\documentclass[useAMS,usenatbib]{mn2e_mod}
\usepackage{graphicx}
\usepackage{amsmath}
\usepackage{amssymb}
\usepackage{deluxetable}
\usepackage{longtable}

\title [Using ML to classify DIBs]{Using machine learning to classify the diffuse interstellar bands}

\author[Baron et al.]
{Dalya Baron$^{1}$\thanks{dalyabaron@mail.tau.ac.il},
Dovi Poznanski$^{1}$\thanks{dovi@tau.ac.il},
Darach Watson$^{2}$,
Yushu Yao$^{3}$,\newauthor
Nick L. J. Cox$^{4,5}$, \& J. Xavier Prochaska$^{6}$\\
$^{1}$School of Physics and Astronomy, Tel-Aviv University, Tel Aviv 69978, Israel.\\
$^{2}$Dark Cosmology Centre, Niels Bohr Institute, University of Copenhagen, Juliane Maries Vej 30, DK-2100 Copenhagen, Denmark.\\
$^{3}$Lawrence Berkeley National Lab, 1 Cyclotron Road, Berkeley CA 94720, USA.\\
$^{4}$Universit$\acute{e}$ de Toulouse, UPS-OMP, IRAP, 31028, Toulouse, France.\\
$^{5}$CNRS, IRAP, 9 Av. colonel Roche, BP 44346, F-31028 Toulouse, France.\\
$^{6}$Department of Astronomy and Astrophysics \& UCO/Lick Observatory, University of California, Santa Cruz, CA 95064, USA.}

\begin{document}

\maketitle

\label{firstpage}
\begin{abstract}

Using over a million and a half extragalactic spectra from the Sloan Digital Sky Survey we study the correlations of the Diffuse Interstellar Bands (DIBs) in the Milky Way. We measure the correlation between DIB strength and dust extinction for 142 DIBs using 24 stacked spectra in the reddening range $\mathrm{E}(B-V) < 0.2$, many more lines than ever studied before. Most of the DIBs do not correlate with dust extinction. However, we find 10 weak and barely studied DIBs with correlations that are higher than 0.7 with dust extinction and confirm the high correlation of additional 5 strong DIBs. Furthermore, we find a pair of DIBs, 5925.9\,\AA\, and 5927.5\,\AA, which exhibits significant negative correlation with dust extinction, indicating that their carrier may be depleted on dust. We use Machine Learning algorithms to divide the DIBs to spectroscopic families based on 250 stacked spectra. By removing the dust dependency we study how DIBs follow their local environment. We thus obtain 6 groups of weak DIBs, 4 of which are tightly associated with $\mathrm{C}_{2}$ or CN absorption lines.

\end{abstract}


\vspace{1cm}
\section{Introduction}\label{s:intro}

The diffuse interstellar bands (DIBs) are unidentified absorption features found in optical and near infrared stellar spectra. Since their discovery in 1919 by \citet{heger22} and the establishment of their interstellar medium (ISM) origin \citep{merrill36, merrill38} they remain an astronomical mystery (see reviews by \citealt{herbig95} and \citealt{sarre06}). Hundreds of DIBs have been found so far and many carriers have been proposed. Polycyclic aromatic hydrocarbons (PAHs) have been suggested by \citet{salama99}, \citet{zhou06} and \citet{leidlmair11}, \citet{kroto85} suggested fullerenes. \citet{Motylewski00}, \citet{Kreowski10} and \citet{Maier11} suggested different hydrocarbon molecules and ions (HC$_5$N$^+$, HC$_4$H$^+$, and linear C$_3$H$_2$ respectively). 

Since the hundreds of known DIBs are unlikely to come from a single carrier (see for example \citealt{herbig91}), searches for correlations between DIBs are carried out with the purpose of revealing sets of DIBs that may come from the same or similar carrier (see reviews by \citealt{herbig95} and \citealt{sarre06}). \citet{weselak04} found that the strength of the $5797.1$\,\AA\, DIB correlates tightly with CH column density. \citet{ehrenfreund02} and \citet{cox08} proved the existence of DIBs and molecular features in the spectra of stars in the Magellanic Clouds and M100, respectively. \citet{kazmierczak09} have found a correlation between the full width half maximum (FWHM) of the $6196$\,\AA\, DIB and the rotational temperature of the $\mathrm{C}_{2}$ molecule. \citet{weselak14} found a correlation between the 4964\,\AA\, DIB and CH and between the 6916\,\AA\, DIB and CH+.

The concept of DIBs families was first described by \citet{krelowski87} and was since developed by others. Some of the studies compute the correlation of equivalent width (EW) or central depth (CD) for pairs of DIBs \citep{chlewicki86,josafatsson87,cami97,moutou99,weselak01,cox05,wszolek06,bryndal07,mccall10,xiang12}, other studies use the correlation with color excess or UV radiation to achieve the task \citep{chlewicki86, krelowski87, josafatsson87, krelowski88, vos11, kos13a}. Moreover, DIBs with complex structure were studied and their structures were compared \citep{jenniskens93}. DIBs strength is the result of a combination of the properties of every physical environment. As the line of sight toward an object usually samples several intervening clouds, studies typically concentrate on observing objects with only one intervening cloud with a known extinction and UV radiation field \citep{cami97}. 

It has been proposed that DIBs spectroscopic families are composed of few strong and a number of weak DIBs \citep{moutou99,fulara00}. 
Recent studies have divided DIBs into three spectroscopic families: (1) DIBs $5797.1$\,\AA\,, $5849.8$\,\AA\,, $6196.2$\,\AA\,, $6379$\,\AA\, and $6613.7$\,\AA\, which have narrow profiles with a substructure, (2) DIBs $5780.6$\,\AA\,, $6284.3$\,\AA\, and $6204.3$\,\AA\, which show no apparent substructure and (3) DIBs $4501.8$\,\AA\,, $5789.1$\,\AA\,, $6353.5$\,\AA\, and $6792.5$\,\AA\, which are weak DIBs for which a substructure is not certain yet \citep{josafatsson87, krelowski87, cami97, porceddu91, moutou99, wszoek03}. DIBs in these families correlate reasonably well among the members in the group but not with those from other families (see \citealt{cox05} for a summary). The pair $6196.2$\,\AA\, and $6613.7$\,\AA\, shows a particularly good correlation of $0.98 \pm 0.18$ and was studied by \citet{moutou99} and \citet{mccall10}, although \citet{galazutdinov02a}, using high resolution and high signal-to-noise ratio (SNR) spectra, argue that the ratio of EW of the two is not always constant.

Most of the DIBs studies are based on observations of a small number of stellar spectra, usually at low Galactic latitude and high extinction. Several studies have also used extragalactic sources, such as nearby galaxies or supernovae, to study the DIBs in the Milky-way (MW) \citep{welty06,cox06b,cordiner08,Cordiner11,van-loon13}.
In recent years there has been an increase in the use of large samples (hundreds and more) of objects in order to draw more general conclusions regarding DIBs  (see for example \citealt{friedman11},\citealt{xiang12}, \citealt{kos13}, \citealt{zasowski14}, \citealt{apellaniz14}, \citealt{lan14a} and \citealt{weselak14}).

\citet{poznanski12} and \citet{baron15} (hereafter paper1) use millions of SDSS spectra of galaxies and quasars to study the properties of the MW Na\,I\,D absorption doublet and DIBs. They show that even though every spectrum has a low SNR, by binning the spectra in large numbers (typically more than 5000) one can detect the strongest DIBs and measure their EW with a 17\%-20\% accuracy. Here we use the central depth (the depth at central wavelength, CD) of DIBs instead of the EW and study the correlation of DIBs and reddening. We show that this method is better suited for our low SNR data and results in correlation uncertainties that are 2--3 times smaller. 

As observing capabilities increase, hundreds of DIBs need to be studied in hundreds or thousands of spectra. This makes grouping and classification via traditional methods an intractable problem. Machine Learning was specifically designed to reveal patterns in large datasets and reduce the dimensionality of such problems. We construct an algorithm that divides the DIBs to spectroscopic families based on their pairwise correlations. The core of the algorithm is the Hierarchical Clustering algorithm, which divides objects to groups based on the distance between them, as defined in some space.

We present the data and our method in section \ref{s:data}, we then study the correlation between DIBs and dust in section \ref{s:flux_corr_red}. We develop an algorithm which computes the pairwise correlation matrix of the DIBs and divides them into spectroscopic families in section \ref{s:DIB-DIB_corrs}, where we also test the algorithm on known atomic and molecular absorption lines. We review and discuss our findings in section \ref{s:conc}. We discuss the extensive simulations we perform to quantify our detection limits and uncertainties in appendices \ref{a:flux-red-sims} and \ref{s:flux_flux_sims}. 

\section{stacking spectra}\label{s:data}

Our reference list of DIBs includes 197 lines which we compile using the catalogs of \citet{jenniskens94}, \citet{jenniskens96}, \citet{krewski95}, \citet{hobbs08} and \citet{jenniskens96}.

The SDSS 9th data release (DR9) comprises of extragalactic spectra that cover an area of more than one forth of the sky, most of which are at high galactic latitudes ($30^{\circ} < b < 90^{\circ}$). Since SDSS spectra are too noisy and have low spectral resolution to show the DIBs we stack them into bins of thousands of spectra. The detailed stacking process is described in paper1. Briefly, we use the galaxies and quasars with redshift $z > 0.005$ from the 9th data release of the SDSS which results in more than 1.5M spectra. The spectra are interpolated to a uniform grid between $3817$\,\AA\, and $9206$\,\AA\, with resolution of $0.1$\,\AA.We then normalize each spectrum using the Savitzky Golay algorithm \citep{savitzky64} and group spectra by different parameters and combine them by calculating the median at every wavelength. 

For the correlation of flux with $\mathrm{E}(B-V)$ in section \ref{s:flux_corr_red} we group the 1.5M spectra by their $\mathrm{E}(B-V)$ values using the dust maps of \citet{schlegel98} as done in paper1. Each of the 24 stacked spectra we obtain is based on stacking about $70\,000$ spectra, and has a SNR greater than 240. For the pairwise correlation between absorption lines in section \ref{s:DIB-DIB_corrs} we group the spectra by their coordinates and their $\mathrm{E}(B-V)$ (see paper1 for details) and result in 250 stacks spectra, each spectrum is a median of $5\,000$ spectra with a SNR near 180. We test different bin sizes for the latter, ranging from $1\,000$ to $40\,000$, and find that the results are consistent. Furthermore, we find that stacking the spectra by coordinates only, regardless of the reddening, yields the same results.

\section{Flux Correlation with E(B-V)}\label{s:flux_corr_red}

\subsection{Method}
DIB strength is usually represented by the EW of the absorption line (see for example \citealt{friedman11} and \citealt{kos13a}) or by its CD \citep{moutou99, weselak01}. Both of the methods have advantages and disadvantages that vary according to the particular study. As shown in paper1, in SDSS spectra, EW can be measured only for $\mathrm{EW} > 5$\,m\AA\footnote{We generally consider a DIB as broad when $\mathrm{FWHM} \gtrsim 2$\AA\, and narrow when $\mathrm{FWHM} \lesssim 1.5$\AA. We refer to strong bands when $\mathrm{EW} \sim 100 \frac{m\AA}{\mathrm{E}(B-V)}$.}, and its use imposes significant uncertainties due to the difficulty to determine the continuum level. Furthermore, the EW is usually measured by integrating over the flux or a fitted profile (Gaussian, Lorentzian or Voight function) between selected endpoints: different methods of EW measurement may yield different results, thus preventing a straightforward comparison between studies. While fitting a profile to uncorrelated blended lines introduces systematic correlation between them and therefore biases their pairwise correlation, simple integration over the flux does not allow the separation of the lines and a blended pair must be treated as a single line. Here we use the flux at the central wavelength of the DIB to represent its strength and measure the correlation between CD and reddening. We show through extensive simulations that this allows us to study weak DIBs ($\mathrm{EW} \ge 6$ m\AA) with the same precision as strong DIBs (typical uncertainty of up to 10\%) and in some cases enables us to separate blended lines.

We use the Pearson correlation coefficient to study the correlation of CD and $\mathrm{E}(B-V)$, 
\begin{equation}\label{eq:1}
{\rho = \frac{cov(x,y)}{{\sigma}_{x}{\sigma}_{y}}.}
\end{equation}
 
We assume that all the DIBs are in the optically thin absorption regime since we use objects at high Galactic latitudes and low dust column densities. This assumption is justified by the linear relations between EW and reddening for the strongest DIBs that were found in paper1, and by the lack of difference between Pearson correlation coefficients and Spearman rank correlation coefficients (which tests for non linear relations) that we measure throughout our entire analysis. 
The flux of an absorption line is given by:\\
\begin{equation}\label{eq:2}
	{I(\lambda) = I_{c}(\lambda) e^{-\tau(\lambda)}}
\end{equation}
Where $I_{c}(\lambda)$ is the flux of the continuum and $\tau$ is the optical depth. In the optically thin regime the optical depth obeys $\tau \ll 1$ and the flux can be expanded:\\
\begin{equation}\label{eq:3}
	{I(\lambda) \cong I_{\lambda}(c)(1-\tau(\lambda))}
\end{equation}
The EW in the optically thin regime is proportional to the optical depth:\\
\begin{equation}\label{eq:4}
{EW = \int_{-\infty}^{\infty} 1 - \frac{I(\lambda)}{I_{c}(\lambda)} d\lambda \sim \tau(\lambda)}
\end{equation}\\
Assuming that the flux of the continuum is constant we obtain:
\begin{equation}\label{eq:5}
{\rho(EW, E(B-V)) = -\rho(I(\lambda_{0}, E(B-V))}
\end{equation}\\
Where $\lambda_{0}$ is the central wavelength.
Lines that get stronger with extinction will have a lower flux and hence a negative correlation between flux and $\mathrm{E}(B-V)$. For clarity we call this simply a `strong correlation' omitting the fact that it is technically negative. Although we do not resolve the DIBs, the CD is linearly related to the EW if the point spread function of the system is approximately Gaussian, which it is.

The normalization of the spectra described in section \ref{s:data} is expected to remove any correlation between the continuum and the reddening and also result in a constant continuum level. On the other hand, the entire stacking process may affect the initial correlation of DIBs and reddening. We therefore simulate synthetic absorption lines with various correlations with reddening to quantify the effects of the entire process on the correlation. The simulations are presented in appendix \ref{a:flux-red-sims}. Using the hundreds of thousands of lines that we simulate, we derive a function that converts between the measured correlation with reddening and the `true' correlation, prior to the stacking process. We find that the ability to detect the correlation of flux with reddening for an isolated absorption line depends only on the slope of the flux--$\mathrm{E}(B-V)$ relation and the width of the line, and that the initial correlation can be recovered for DIBs with a slope greater than $10^{-2.9} \frac{1}{mag}$ and FWHM greater than $0.5$\,\AA. We find that the initial correlation of a blended pair of lines depends on both of the measured correlations and can be recovered above the same slope threshold. Since we lack the resolution to measure the width of the DIBs we use the average width value from DIB catalogs to exclude DIBs which have widths that are below our threshold. However, we measure the correlation of DIBs with borderline widths values (e.g the measured FWHM of the $6196$\,\AA\, DIB is in the range $0.2-0.8$\,\AA\, according to different studies) and these should be treated with caution. In this case, low measured correlation can indicate a non detection. We mark the borderline DIBs in the online material.

We use the 24 spectra, stacked using the \citet{schlegel98} maps, as described in section \ref{s:data}, and measure the correlation coefficient between flux and reddening for every wavelength in our wavelength grid. The result is a continuous correlation spectrum. 

\subsection{Results}\label{s:dibs_corr_red}

\begin{figure*}
\includegraphics[width=0.99\textwidth]{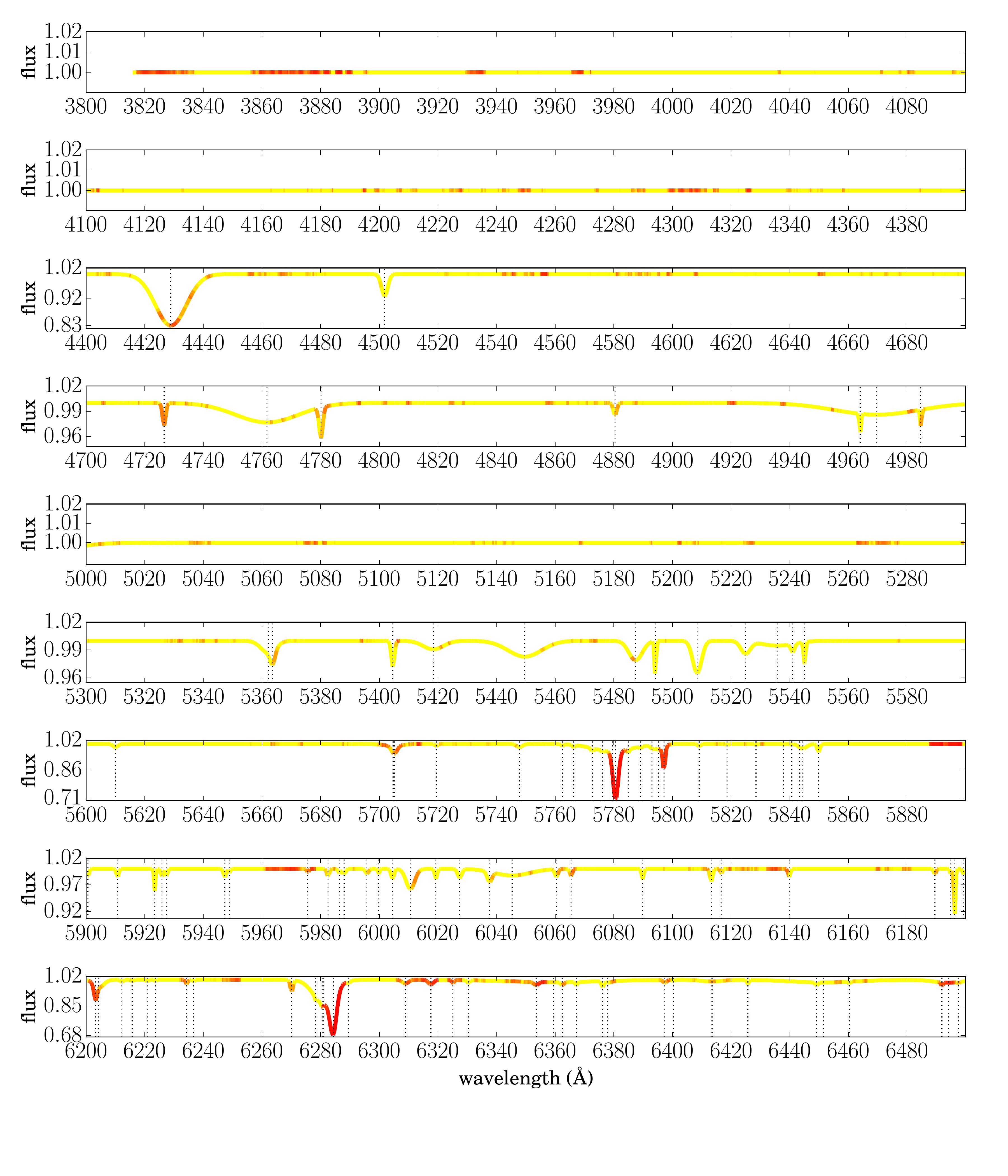}
\caption{Correlation spectrum for the wavelength range \,$3800-6500$\,\AA. The spectrum is a simulated spectrum based on the DIBs catalog and is colored by the correlation coefficient between flux and E(B-V), red for strong (negative; since in absorption) correlation (-1) and yellow for a correlation that is close to zero. The colors are clipped to correlation values of (0.5,1) for clarity.}\label{f:flux_ext_corr_1_1}
\end{figure*}

\begin{figure*}
\includegraphics[width=0.99\textwidth]{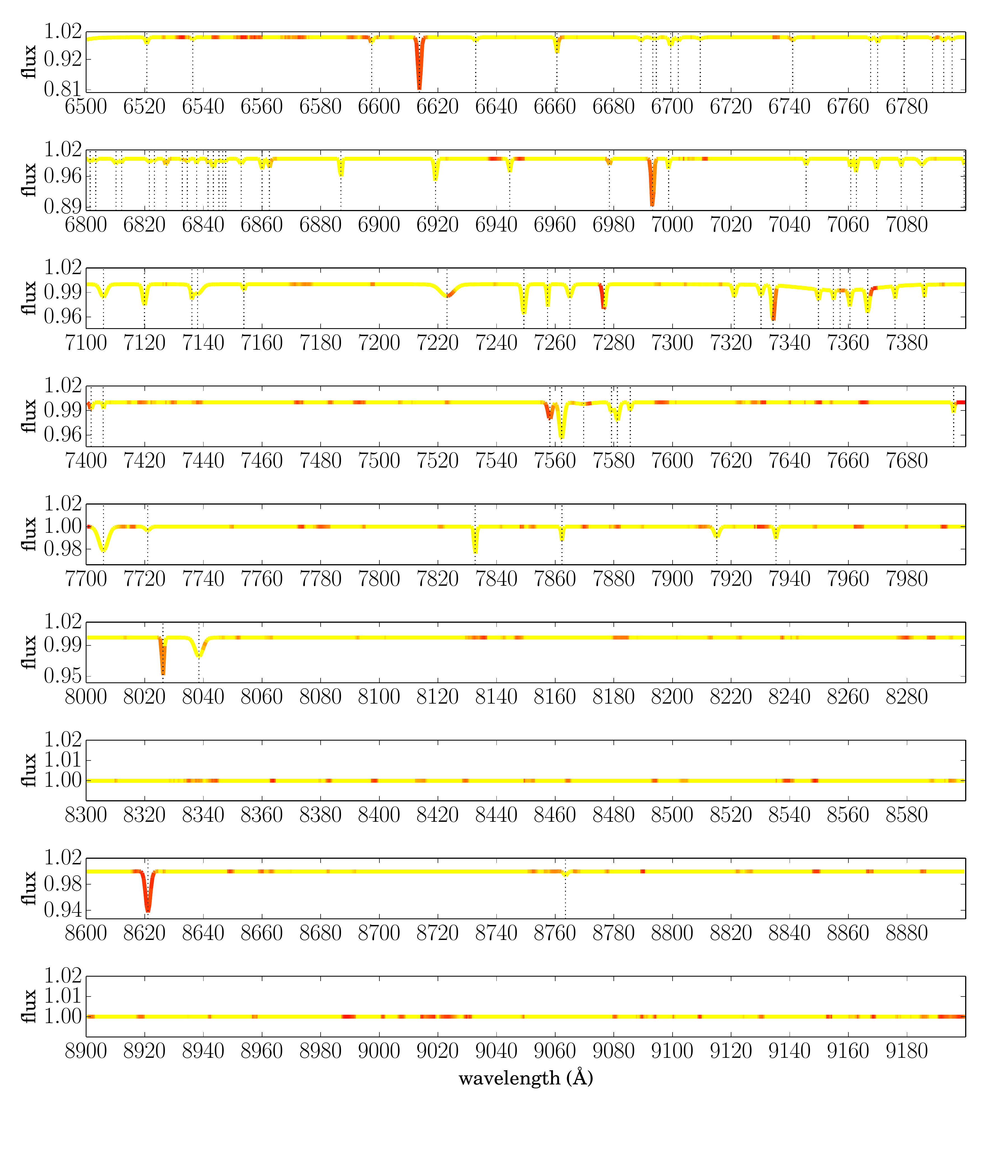}
\caption{Same as Figure \ref{f:flux_ext_corr_1_1}, for the wavelength range \,$6500-9200$\,\AA.}\label{f:flux_ext_corr_2_1}
\end{figure*}

\begin{figure*}
\includegraphics[width=0.99\textwidth]{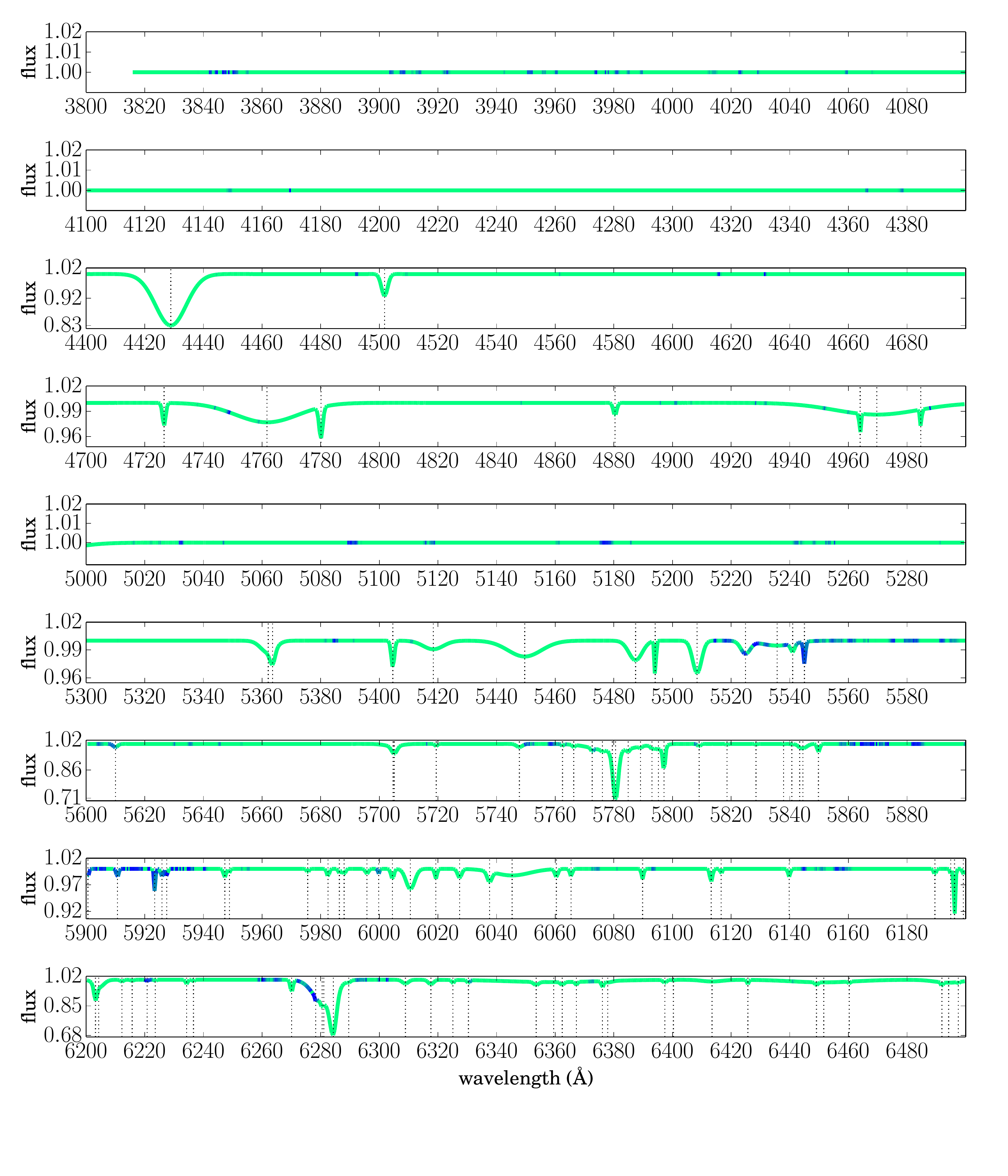}
\caption{Correlation spectrum for the wavelength range \,$3800-6500$\,\AA. The spectrum is a simulated spectrum based on the DIBs catalog and is colored by the correlation coefficient between flux and E(B-V), blue for strong correlation (1) and green for a correlation that is close to zero. The colors are clipped to correlation values of (-0.5,-1) for clarity.}\label{f:flux_ext_corr_1_1_pos}
\end{figure*}

\begin{figure*}
\includegraphics[width=0.99\textwidth]{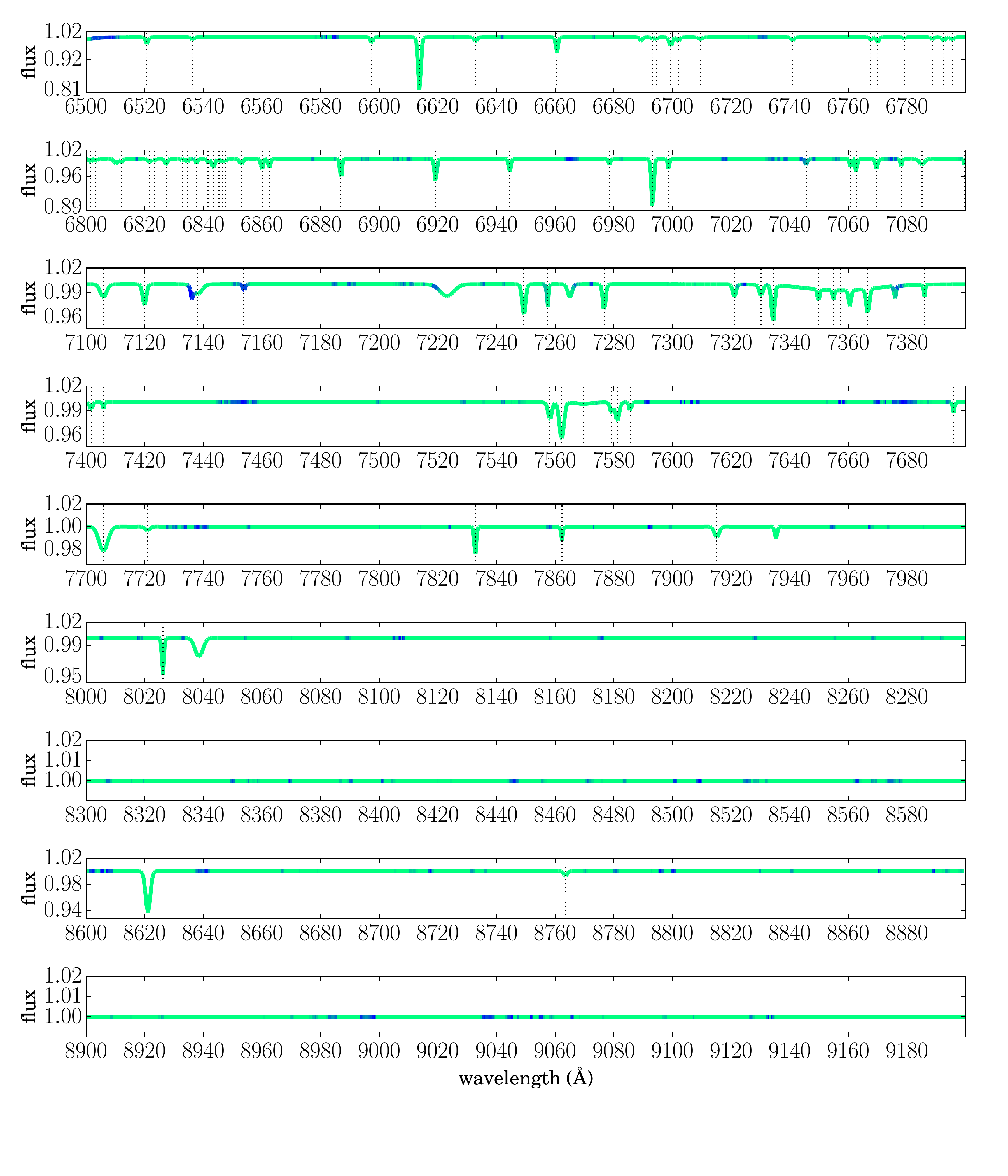}
\caption{Same as Figure \ref{f:flux_ext_corr_1_1_pos}, for the wavelength range \,$6500-9200$\,\AA.}\label{f:flux_ext_corr_2_1_pos}
\end{figure*}

For presentation, we color a synthetic spectrum based on the DIB catalog according to the correlation we measure. We clip the colors in figures \ref{f:flux_ext_corr_1_1}--\ref{f:flux_ext_corr_2_1_pos} to strong correlation values, for clarity. Red regions in figures \ref{f:flux_ext_corr_1_1} and \ref{f:flux_ext_corr_2_1} mark high correlation of flux with color excess (negative; since in absorption), while yellow coloring implies little to no correlation. Blue regions in figures \ref{f:flux_ext_corr_1_1_pos} and \ref{f:flux_ext_corr_2_1_pos} mark high correlation (positive), while green coloring implies little to no correlation. Clearly some lines show strong correlation (e.g. $5780.6$\,\AA\, and $5797.1$\,\AA\,) while others, often as strong, do not (e.g. $4428$\,\AA). Furthermore, one can see the correlation with dust of interstellar absorption lines which are not DIBs.   

Our simulations show that we can robustly recover the true correlation only above a threshold in the slope of the flux-reddening relation. 154 DIBs are above the threshold, thus their correlation with reddening can in principle be measured. Of the 154 DIBs, 132 are isolated and 22 are blended in 11 pairs. 4 pairs are discarded since they exhibit correlations with opposite signs and 2 pairs are discarded since they exhibit correlations that are lower than 0.5 (see appendix \ref{a:flux-red-sims} for details), thus 142 DIBs remain.
A sample of the correlations we measure is given in table \ref{t:corr_red_strong}, the full table can be found in the online version of this paper. The correlation uncertainties we deduce for the isolated DIBs vary from an average of 5\% for high correlations (0.8)  up to 15\% for low correlations above 0.3. We find in paper1 that the uncertainty of the EW for the strongest DIB, $5780.6$\,\AA\, is 17\% (and larger for smaller DIBs) and the uncertainty in the correlation is of order 20\%. These results are therefore 2--3 times better than in paper1.

Of the 142 DIBs, only 15 isolated DIBs and 2 pairs of blended DIBs have a statistically significant correlation with $\mathrm{E}(B-V)$ that is higher than 0.7. While the high correlation of the DIBs: $5705$\,\AA\, (pair), $5780$\,\AA\,(pair), $5797.1$\,\AA, $6284.3$\,\AA, $6613.7$\,\AA\, and $8621.1$\,\AA\, is a confirmation of what was already established in previous studies (see for example \citealt{friedman11}, \citealt{kos13a} and \citealt{kos13}), we find high correlations with dust extinction for much weaker, rarely studied DIBs: $5975.6$\,\AA, $6308.9$\,\AA, $6317.6$\,\AA, $6325.1$\,\AA, $6353.5$\,\AA, $6362.4$\,\AA, $6491.9$\,\AA, $6494.2$\,\AA, $6993.2$\,\AA, $7223.1$\,\AA\, and $7558.2$\,\AA. 

Moreover, we find a blended pair of DIBs, 5925.9\,\AA\, and 5927.5\,\AA, which exhibit a high positive correlation between their flux and $\mathrm{E}(B-V)$, i.e., negative correlation between their strength and dust extinction. These DIBs are the only ones out of the 142 bands that exhibit such strong negative correlation with dust extinction, they show correlations that are consistent with being similar and they are highly blended, thus they are likely to originate from the same carrier, perhaps one that is easily depleted on dust.

We find that the DIBs $4428$\,\AA, $5512$\,\AA, $5850$\,\AA, and $6445$\,\AA\, are absent in our data and the DIBs $4762$\,\AA, $4964$\,\AA, $5487$\,\AA, $6090$\,\AA, and $6379$\,\AA\, have a low correlation with dust (lower than 0.5). The above are among the strongest and/or most studied DIBs, most of which exhibit medium to high correlations with dust in the Galactic plane (see for example: \citealt{friedman11} and \citealt{kos13a}). Since we measure strong and significant correlations with dust extinction for other DIBs with similar strengths and correlations we conclude that the above indeed behave differently in our data.

Established by \citet{krelowski87} and later studied and adjusted by a number of studies \citep{chlewicki86, krelowski88,krewski95,cami97,weselak01,kos13a}, the difference in DIBs behavior in different environments has led to the suggestion that UV radiation leads to the weakening of some DIBs while not affecting other DIBs or even strengthening them. The major classes that are said to exist are $\sigma$ (not UV shielded) and $\zeta$ (UV shielded) clouds. This partition was made using the strengths of the $5780.6$\,\AA\, and $5797.1$\,\AA\, DIBs. While the $5780.6$\,\AA\, DIB appear stronger where the UV radiation is stronger ($\sigma$ type), the $5797.1$\,\AA\, appears stronger in UV shielded environments ($\zeta$ type). Following this, \citet{kos13a} divide their sight lines into $\sigma$ and $\zeta$ type clouds based on the ratio of the $5797$\,\AA\, and $5780$\,\AA\, DIBs strength and find that this partition shows at least two different types of DIBs: type I DIBs where the behavior of the DIB EW and reddening is similar for $\sigma$ and $\zeta$ sight lines ($5705$\,\AA, $5780$\,\AA, $6196$\,\AA, $6202$\,\AA\, and $6270$\,\AA) and type II DIBs where it differs significantly ($4964$\,\AA, $5797$\,\AA, $5850$\,\AA, $6090$\,\AA, $6379$\,\AA, $6613$\,\AA\, and $6660$\,\AA). The DIBs we find to be absent or to have low correlation with $\mathrm{E}(B-V)$ are inconsistent with these groups and with similar groups that were established by other studies using the $\sigma$/$\zeta$ partition i.e., some of the type I DIBs from \citet{kos13a} are absent in our data, some show low correlation with dust and some show the opposite. The same applies to type II DIBs.

\citet{mcintosh93} studied the strength of the DIBs $4428$\,\AA, $5780.6$\,\AA\ and $5797.1$\,\AA\ as a function of Galactic latitude. They found that the strength of $4428$\,\AA\ relative to the others is greatest at low latitude and decreases with increasing latitude whereas the strength of $5797.1$\,\AA\ is greatest at high latitude. \citet{van-loon13} find that the growth of the EW of the $4428.1$\,\AA\, appears to slow down as the EW of the $5780$\,\AA\, DIB grows. We show in paper1 that the $4428$\,\AA\, is absent in our spectra while the $5780.6$\,\AA\ and $5797.1$\,\AA\ DIBs tend to have higher EW per reddening unit compared to the more UV shielded Galactic plane, in agreement with the findings of \citet{mcintosh93} and \citet{van-loon13}. We therefore suggest that the absence and weakening of some of the strongest, most studied, DIBs is not due to UV radiation but highly connected to the unique environment we probe in this study. Our results in paper1 reinforce the findings of \citet{mcintosh93} and our current results add more DIBs that follow the same patterns. The group $4428$\,\AA, $5512$\,\AA, $5850$\,\AA, and $6445$\,\AA\, tends to disappear in high Galactic latitudes while the group $4762$\,\AA, $4964$\,\AA, $5487$\,\AA, $6090$\,\AA\ and $6379$\,\AA\, exhibit almost no correlation with dust extinction ($0 - 0.4$).

\begin{table}
\caption{DIB correlation with reddening} 
\centering 
\begin{tabular}{lc}
\hline\hline 
Wavelength\tablenotemark{*} [\AA] & Deduced correlation\\ [0.5ex] 

\hline 
5704.7\tablenotemark{*} &  $0.99\pm0.50$ \\ 
5705.1\tablenotemark{*} & $0.99\pm0.54$ \\ 
5779.5\tablenotemark{*} & $1.0\pm0.055$ \\ 
5780.6\tablenotemark{*} & $1.0\pm0.055$ \\
5797.1 & $0.87\pm0.02$ \\
5925.9\tablenotemark{*} & $-0.9991\pm0.0004$ \\ 
5927.5\tablenotemark{*} & $-0.9997\pm0.0002$ \\
5975.6 & $0.72\pm0.03$ \\
6203.2\tablenotemark{*} & $0.99\pm0.30$ \\
6204.3\tablenotemark{*} & $0.99\pm0.31$ \\
6280.5\tablenotemark{*} &  $0.99\pm0.11$ \\
6281.1\tablenotemark{*} & $0.99\pm0.12$ \\
6284.3 & $0.976\pm0.029$ \\
6308.9 & $0.762\pm0.025$ \\
6317.6 & $0.841\pm0.022$ \\
6325.1 & $0.713\pm0.025$ \\
6353.5 & $0.839\pm0.023$ \\
6362.4 & $0.702\pm0.025$ \\
6491.9 & $0.808\pm0.025$ \\
6494.2 & $0.904\pm0.024$ \\
6613.7 & $0.801\pm0.025$ \\
6993.2 & $0.704\pm0.025$ \\
7223.1 & $0.706\pm0.025$ \\
7558.2 & $0.777\pm0.026$ \\
8621.1 & $0.841\pm0.022$ \\

 
\hline 
\tablenotetext{*}{We mark DIBs that are part of a blended pair.}
\end{tabular} 
\\[10pt]
Correlation between CD and reddening for the DIBs that exhibit a strong correlation. The full table is available in the online version of this paper. 
\label{t:corr_red_strong}
\end{table}

\section{Clustering lines}\label{s:DIB-DIB_corrs}

As we show above, while some DIBs correlate fairly well with color excess, many more do not, and there is substantial intrinsic scatter (see also \citealt{vavloon09}, \citealt{friedman11}, \citealt{kos13a}, \citealt{van-loon13}, and paper1). Furthermore, recent studies have measured DIBs strength in low column densities environments and found that the scale height of the DIBs carriers may be more extended than the distribution of dust (paper1 for $5780.6$\,\AA\, and $5797.1$\,\AA\, DIBs and \citealt{kos14} for $8621.1$\,\AA).

Following these findings we wish to compute the pairwise correlation of DIBs after removing the correlation with dust, basically using dust as a tracer for gas column density, which is a nuisance parameter. The result of the process is a 197X197 matrix that represents the pairwise correlation of all the DIBs. We then wish to divide the DIBs to spectroscopic families based on this correlation matrix. The algorithm, which takes spectra as an input and returns groups of lines (DIBs or others) as an output, is described in depth in section \ref{s:algo}. We test the algorithm, find its detection limits and characterize its precision and accuracy in appendix \ref{s:flux_flux_sims}. We further test the algorithm on known atomic and molecular absorption lines in section \ref{s:mol_groups}, and discuss the results for DIBs in section \ref{s:dibs_groups}. 

\subsection{Algorithm}\label{s:algo}

For clarity, we describe the algorithm as it works with an input of 250 binned spectra, each spectrum has a median extinction value and 8000 wavelength elements from 4000\,\AA\, to 8000\,\AA\, with a resolution of 0.1\,\AA.
In order to compute the correlation between pairs of fluxes one must first remove the correlation between flux and reddening, which we do by computing the partial correlation:
\begin{equation}\label{eq:7}
{{\rho}_{f_{1}f_{2},A_{v}} = \frac{{\rho}_{f_{1}f_{2}}-{\rho}_{f_{1}A_{v}}{\rho}_{A_{v}f_{2}}}{{\sqrt{1-{\rho}_{f_{1}A_{v}}^2}}{\sqrt{1-{\rho}_{A_{v}f_{1}}^2}}}}
\end{equation}
Where the extinction $A_{v}$ is the nuisance parameter, $f_{1}$ and $f_{2}$ is the pair of fluxes we wish to compute partial correlation for, and $\rho_{ab}$ is the Pearson Correlation Coefficient between the $a$ and $b$ variables.

Computing the partial correlation between every individual pair of fluxes in the wavelength grid returns a 8000 X 8000 correlation matrix. The [$i$,$j$] element of the correlation matrix represents the partial correlation between the fluxes in the $i${th} and $j$th wavelength indices. Therefore, the correlation matrix is symmetric, positive-definite, and its primary diagonal is equal to one. 

\begin{figure}
\includegraphics[width=3.25in]{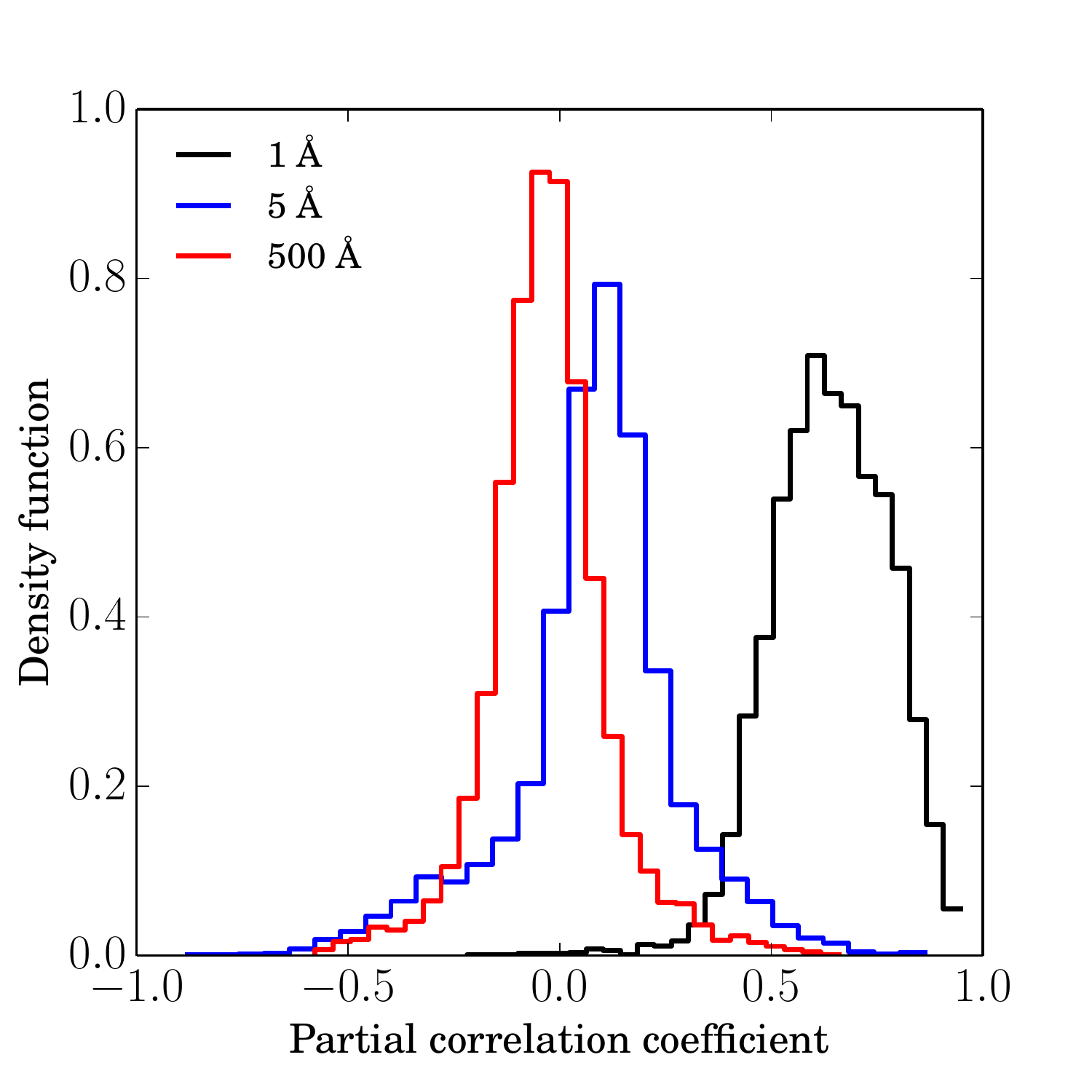}
\caption{Correlation distribution as a function of distance in wavelength. The flux-flux correlation of wavelengths that are separated by 1\,\AA\, (black) show a median correlation of 0.65, wavelength that are separated by 5\,\AA\, (blue) show a median correlation of 0.1 and wavelengths that are separated by 500\,\AA\, (red) have a median correlation of 0.}\label{f:diag_hist}
\end{figure}

\begin{figure}
\includegraphics[width=3.25in]{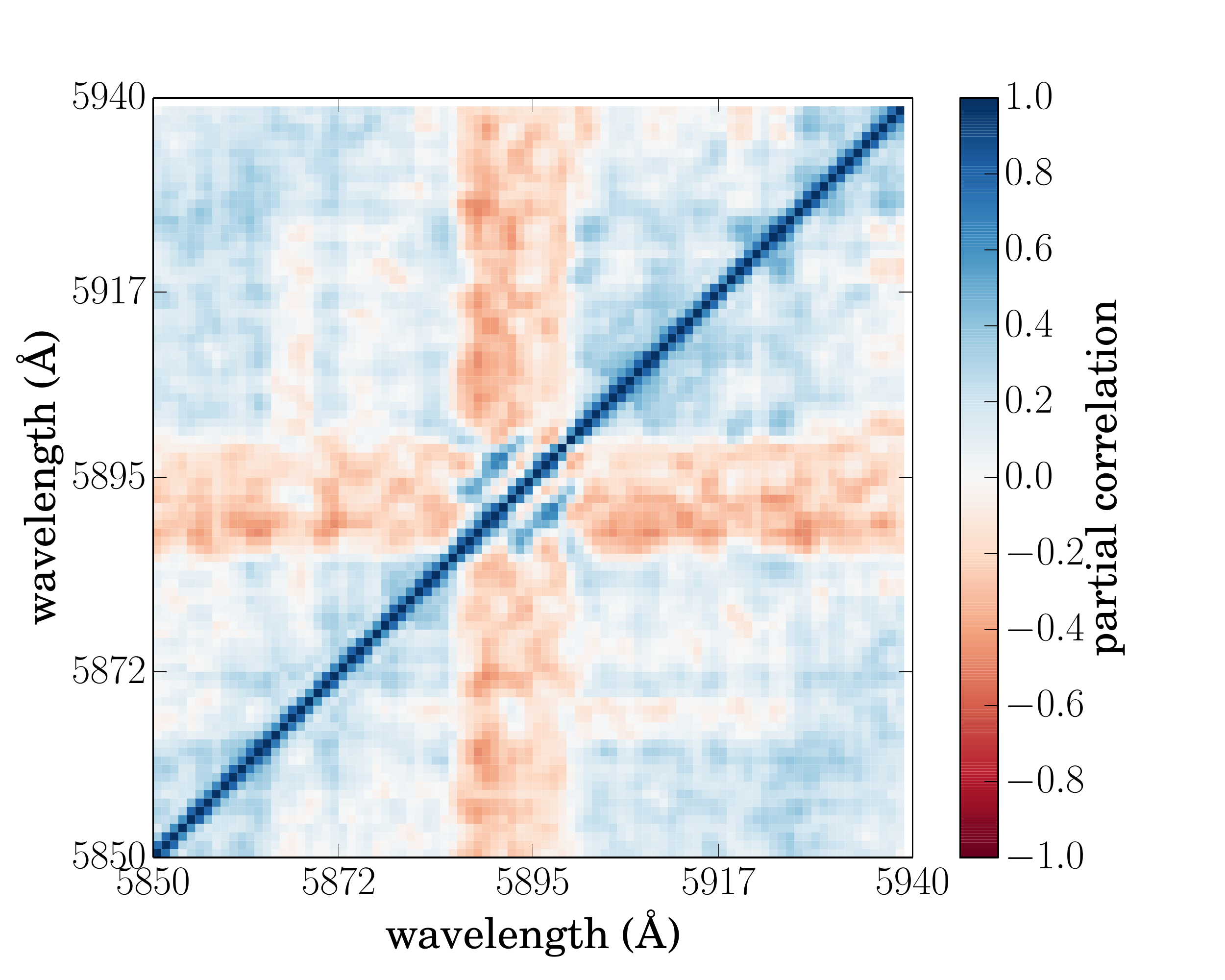}
\includegraphics[width=3.25in]{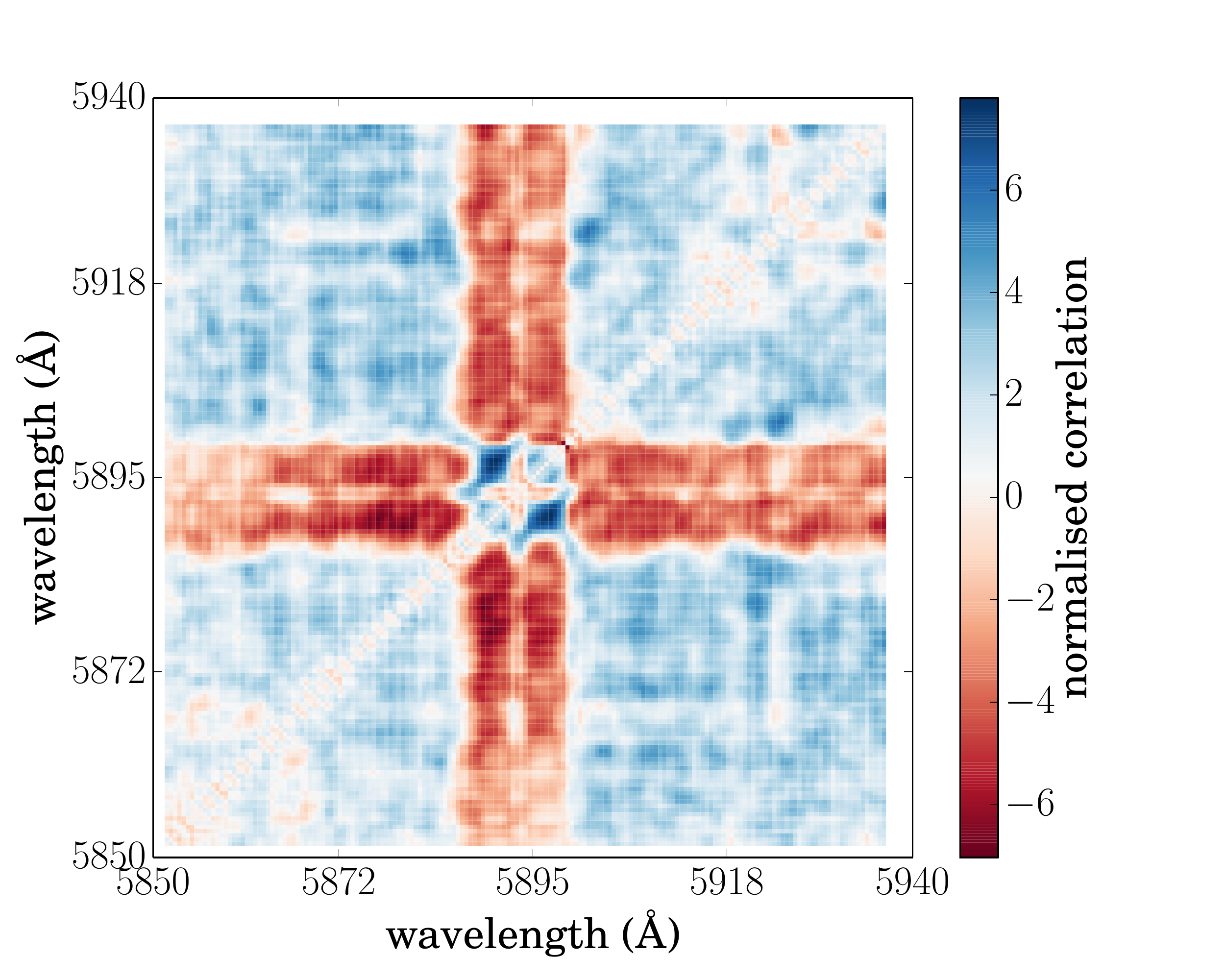}
\caption{Partial correlation matrix cutout around the Na\,I\,D wavelengths (top) and the normalized correlation matrix (bottom).}\label{f:Naid_corr}
\end{figure}

Diagonals which are far away from the primary diagonal follow a Normal distribution with zero mean. This is expected since randomly chosen distant wavelengths should not be correlated. Diagonals which are close to the primary diagonal measure the flux-flux correlation of nearby wavelengths and their mean is shifted towards 1 as the diagonal gets closer to the primary diagonal. This bias is the result of our limited resolution and some systematic effects that are due to the calibration and analysis process. We wish to remove it. Figure \ref{f:diag_hist} presents a sample of correlation distributions for different diagonals. In order to remove the spurious correlation that is caused by wavelength proximity we normalize each element in the correlation matrix in the following way:
\begin{equation}\label{eq:8}
{\rho_{norm} = \frac{\rho - \mu_{k}}{\hat{\sigma_{k}}}}
\end{equation}
Where $\mu_{k}$ is the median correlation for the $k$th diagonal and $\hat{\sigma}_{k}$ is the median absolute deviation (MAD) of the $k$th diagonal. We thus obtain a 8000 X 8000 normalized correlation matrix in which the [$i$, $j$] element represent the deviation in $\sigma$ units from the median correlation. We consider $|\rho_{norm}| > 3\sigma$ as high correlation (see discussion below). We present a cutout of the partial correlation matrix and normalized correlation matrix around the Na\,I\,D wavelengths in figure \ref{f:Naid_corr}. Despite the proximity of the doublet lines one can see that the strong positive correlation between them remains in the normalized matrix. It is also apparent that the correlated lines produce a secondary effect of strong negative correlation that fades away as the element is farther away from the Na\,I\,D wavelength. This does not affect our final results.

\begin{figure}
\includegraphics[width=3.25in]{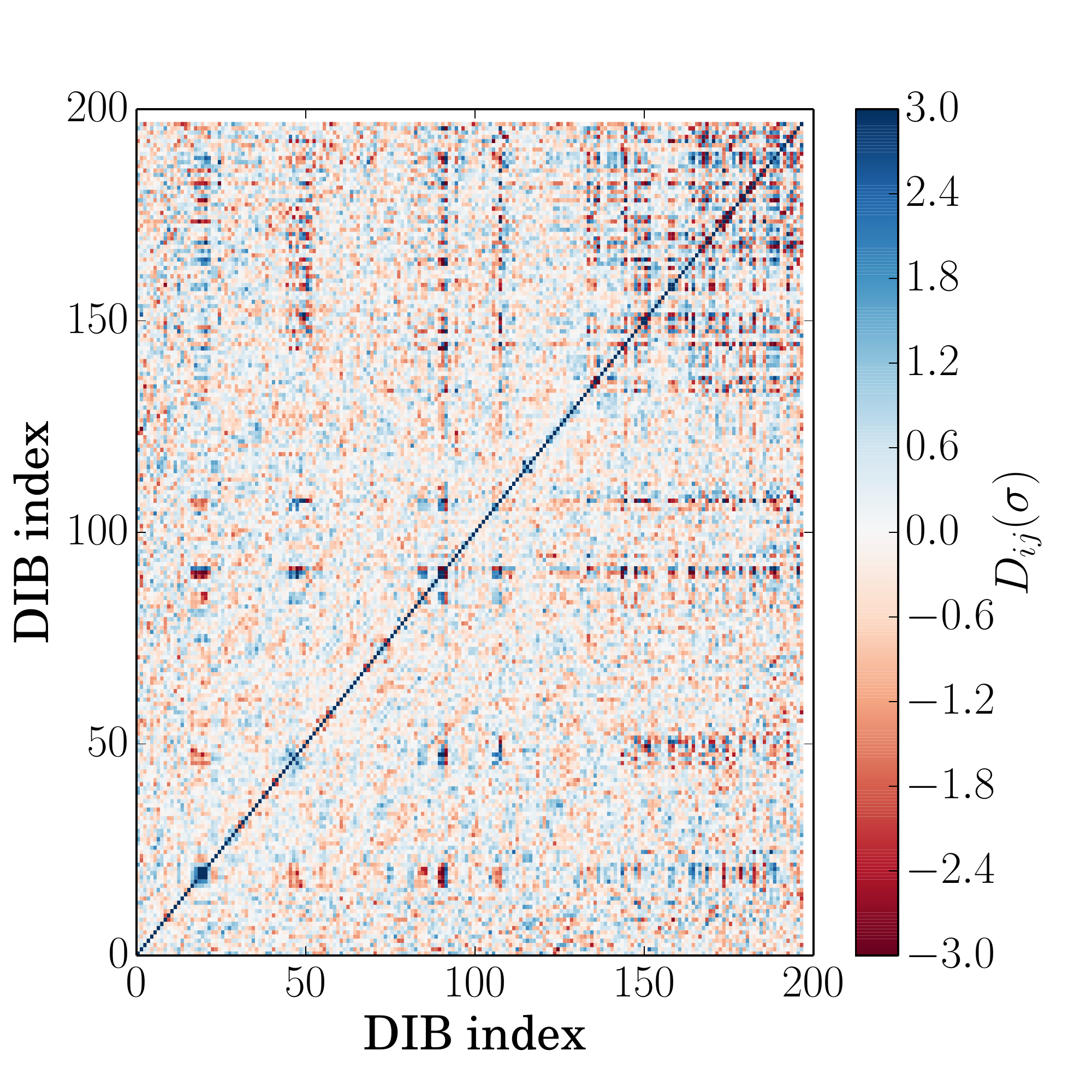}
\caption{197 by 197 correlation matrix for the DIBs. Every cell represents the partial (normalized) correlation coefficient between a pair of DIBs after the correlation with E(B-V) is removed. Blue (red) cells represent strong positive (negative) correlation between lines. We clip the correlations in the figure to [$-3\sigma$, $3\sigma$] for clarity, from the original range of  [$-8\sigma$, $8\sigma$]. The wavelengths that correspond to the DIBs indices are available online.}\label{f:flux-flux-corr}
\end{figure}

We then reduce the normalized correlation matrix to a matrix that contains only the elements of the pairwise correlation between absorption lines we wish to study, using their central wavelength. We tested several methods to obtain the reduced matrix. The simplest is taking the elements [$i$,$j$] that match the $i$th and $j$th central wavelengths. We also tried integrating over the wavelength range of a given line using its measured profile as a weight function. We find that the latter method smears the resulting correlation, some of the smearing is caused by the secondary effect of the negative correlation. For the NaID doublet, for example, we measure a correlation significance of $4.3\sigma$ while in the bottom of figure \ref{f:Naid_corr} the significance at the central wavelengths is $7.8\sigma$. We therefore use the exact wavelengths to extract the correlation element and obtain the reduced matrix. 
Figure \ref{f:flux-flux-corr} presents the 197 X 197 normalized correlation matrix that contains the pairwise correlation of the DIBs. 
The correlation matrix contains $19\,306$ unique elements, which prevents us from dividing the DIBs into groups manually. We use a Hierarchical Clustering algorithm to achieve this goal.

Hierarchical Clustering is a Machine Learning algorithm that seeks to build a hierarchy of clusters (i.e., groups) based on the distance between objects. We use the agglomerative type of clustering, the bottom-up approach, in which each object (absorption line in our case) starts as a one sized cluster, and pairs of clusters are merged according to their distance as one moves up the hierarchy. In order to convert correlations between lines into distances, we perform the following linear transformation on the normalized correlation matrix, such that a strong positive correlation transforms to a short distance, and a strong negative correlation transforms to a long distance:

\begin{equation}\label{eq:9}
{{D}_{ij} = max({R}_{ij}) - {R}_{ij}}
\end{equation}

Where $R_{ij}$ is the normalized and reduced correlation matrix that is shown in figure \ref{f:flux-flux-corr}. The transformation yields the matrix $D_{ij}$ which represents the pairwise distance between DIBs. 

Every DIB starts as a one-sized cluster. We choose the pair with the closest distance and merge it into a two-sized cluster. We then define the distance between the newly formed cluster to the rest of the clusters (so far they all contain only one DIB) as the maximal distance between the members in the clusters, i.e., the lowest correlation between the members of the clusters. We repeat these steps until all the DIBs are clustered.

This method is called complete linkage clustering, in which new distances are defined as the maximal distance between all the members of the two new-formed clusters. The clusters that are formed using this method are the worst case result since we group the DIBs by the lowest correlation of the bunch.
Our tests indicate that other choices for the distance assigning: single linkage (minimal distance, maximal correlation), average linkage, and weighted linkage produce almost the same clustering and very similar groups.

\begin{figure}
\includegraphics[width=3in]{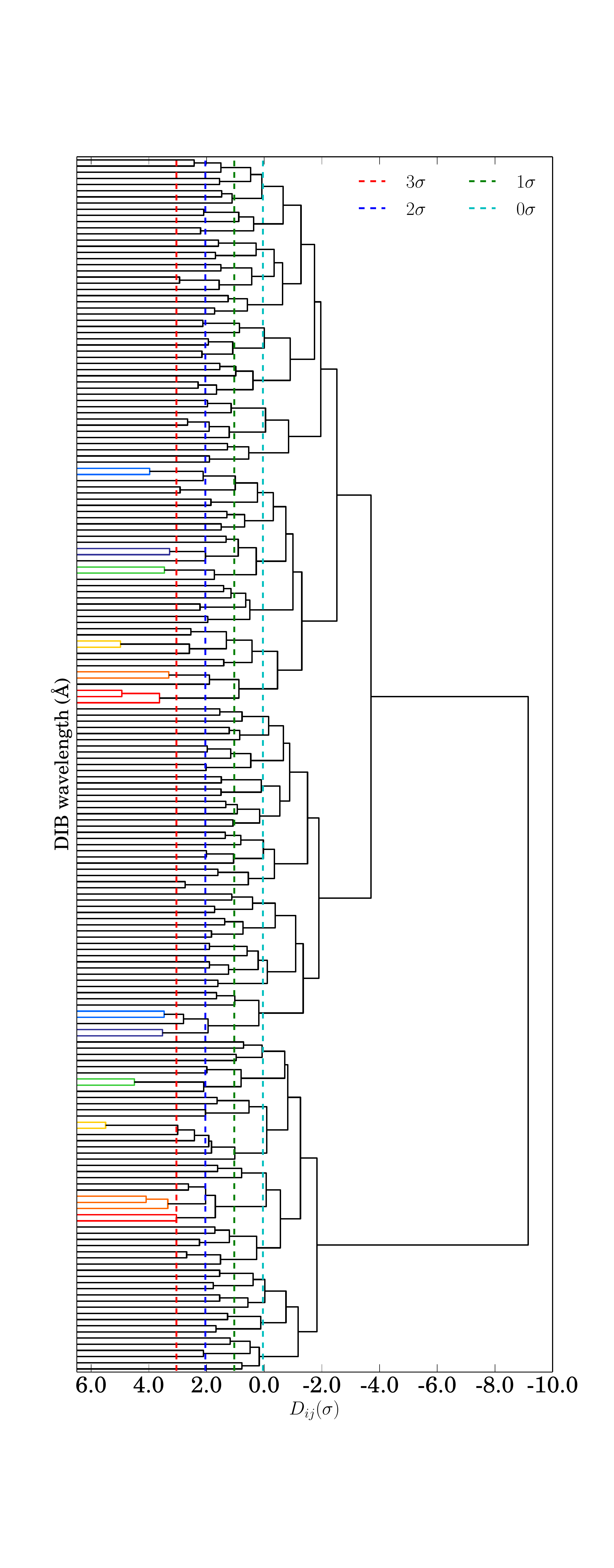}
\caption{DIBs clustering by the partial correlation between their flux. The colored clusters represent the groups for which the pairwise correlation is higher than $3\sigma$. A larger version of this figure with DIB wavelength labels can be found in the online material.}\label{f:cluster_results_dibs}
\end{figure}

The result of the clustering for the correlation matrix in figure \ref{f:flux-flux-corr} is presented in figure \ref{f:cluster_results_dibs}. One can see that the result of the clustering contains colored branches that are sub-clusters in which all the members have a pairwise correlation of more than $3\sigma$. We define the sub-clusters that fall beneath this threshold as spectroscopic families. A larger version of this figure with DIB wavelength labels can be found in the online material.

The algorithm discussed above consists of several phases, each of the phases can be implemented in various ways. We perform extensive simulations throughout the development of the algorithm in order to test every step and find the best implementation to it. The simulations are discussed thoroughly in appendix \ref{s:flux_flux_sims}, by injecting synthetic absorption lines with correlations among them to the raw spectra we form synthetic spectroscopic groups. We compare these groups to the groups we obtain from the algorithm and statistically infer the probability of the resulting group to be a true-positive (TP) group or a false-positive (FP) group. We find that the probability of a resulting spectroscopic group to be a TP or FP depends on the correlation of the lines in it with reddening and on their slopes of the flux-$\mathrm{E}(B-V)$ relation. Throughout the next section we assign probabilities for a group to be TP or FP based on these simulations.

\subsection{Atomic and molecular lines}\label{s:mol_groups}

Known molecular and atomic absorption lines can be used as an additional test to our clustering algorithm. Absorption lines of the same molecule or atom should be highly correlated and be clustered as one spectroscopic family (though different levels may be populated in different environments, complicating the picture). We use a list of 114 atomic and molecular absorption lines (Cox private communication) that contains the central wavelengths of CH, CH+, NH, OH, OH+, CN and $\mathrm{C}_2$ molecules and Al\,I, Ca\,I, Ca\,II, Cr\,I, Fe\,I, He\,I, K\,I, Li\,I, Na\,I, Rb\,I, and Ti\,II atoms. Following the procedure described above, we obtain a 114 X 114 normalized correlation matrix for these lines. We note that this includes lines that are too weak to be detected in our spectra, which is a good test of the applicability of our method to DIBs that are also not all detectable.

\begin{figure*}
\includegraphics[width=0.8\textwidth]{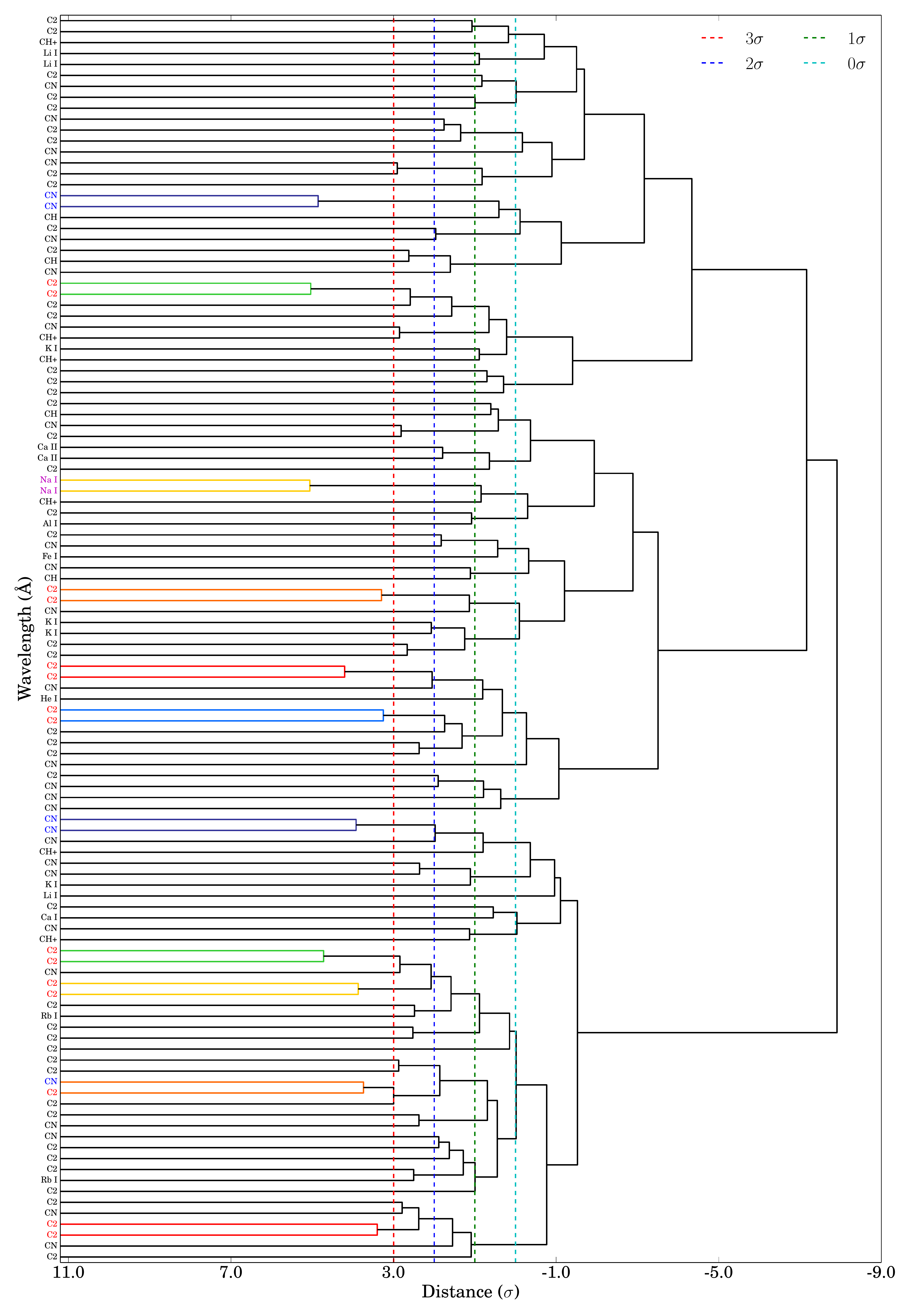}
\caption{Clustering result for atoms and molecules.}\label{f:clustering_atoms}
\end{figure*}

We present the clustering result for atoms and molecules in figure \ref{f:clustering_atoms}. 11 clusters of lines form, each containing 2 absorption lines with a pairwise correlation that is higher than $3\sigma$. Each significant cluster is assigned a color in the figure. Different atoms and molecules have different label colors. We expect and see that clusters are almost always populated by a single species. 10 out of the 11 formed groups contain a pair of absorption lines that belong to the same molecule or atom while only one group contains a pair of lines that belong to different carriers. 

\begin{table}
\setlength{\tabcolsep}{3pt}
\caption{Atoms and molecules groups for $3\sigma$ threshold} 

\begin{tabular}{lcccc} 
\hline \hline 

Group ID & WLs[\AA] & Carrier & P(0)\tablenotemark{a} & P(TP)\tablenotemark{a} \\ [0.5ex]

\hline

1 & 7720.7, 8782.3 & C2 & 0.23 & 0.64 \\
2 & 8853.0, 9142.8 & C2, CN & 0.02 & 0.4 \\
3 & 7722.1, 7731.0 & C2 & 0.05 & 0.42 \\
4 & 8798.5, 8804.5 & C2 & 0.12 & 0.55 \\
5 & 6929.1, 9183.2 & C2 & 0.13 & 0.52 \\
6 & 8762.1, 8768.6 & C2 & 0.17 & 0.68 \\
7 & 8761.2, 8767.8 & C2 & 0.4 & 0.77 \\
8 & 8757.7, 8766.0 & C2 & 0.36 & 0.75 \\
9 & 5890.0, 5895.9 & NaID & 0.08 & 0.73 \\
10 & 7753.1, 7762.6 & C2 & 0.1 & 0.62 \\
11 & 9186.9, 9190.1 & CN & 0.11 & 0.56 \\

\hline 
\hspace{-0.1cm}
\tablenotetext{a}{Probability for perfect recovery.}
\tablenotetext{b}{Probability for TRUE POSITIVE = P(0) + P(1)}
\end{tabular} 
\label{t:3sig_groups}
\end{table}

We list in table \ref{t:3sig_groups} the 11 groups, the absorption lines they contain and their probabilities to be a perfectly recovered group ,P(0), and a TP group. The groups probability to be a FP group is 1-P(TP). We are able to recover 7 groups of $\mathrm{C}_2$ absorption lines, 2 groups of 12CN absorption lines and the group of the Na\,I\,D atom. We find that 10 of the groups have a probability to be a TP group that is higher than 0.5 while 2 of the groups present probabilities of 0.4 and 0.42 to be a TP group. Group 2, with the lowest probability in the bunch, is also the group which contains absorption lines from different carriers. 

A few conclusions are apparent. First, the clustering works well, in the sense that 10/11 groups are correct. Second, very few lines cluster, most remain single. We discuss this point further in section \ref{s:opt}.  Last, we see that our clustering algorithm is consistent and that the simulations we perform are representative of the real lines, since the probabilities point  well towards our successes and failures. The fact that P(0) is typically very small reflects that the algorithm only recovers partial groups, as is apparent for example for $\mathrm{C}_2$ lines that are scattered among multiple groups. Examining the 22 clustered lines in our spectra, we see that they can be seen by eye, while most unclustered lines are not visible. 

\subsection{DIBs results}\label{s:dibs_groups}

The DIBs 197 X 197 clustering normalized partial correlation matrix is shown in figure \ref{f:flux-flux-corr} and the clustering result in figure \ref{f:cluster_results_dibs}. We obtain 12 DIB groups, 10 of which contain two absorption lines while two of them contain three absorption lines. We compute the probability of every group to be a TP group, and list them in table \ref{t:3sig_groups_dibs}. 

\begin{table}
\setlength{\tabcolsep}{3pt}
\caption{DIBs groups for $3\sigma$ threshold} 

\begin{tabular}{lccc} 
\hline \hline 

Group ID & DIBs[\AA] & P(0)\tablenotemark{a} & P(TP)\tablenotemark{a} \\ [0.5ex]

\hline
3 & 6317.6, 6325.1 & 0.28 & 0.74 \\
2 & 6497.5, 6862.5, 7569.7 & 0.27 & 0.7 \\
4 & 7276.7, 8026.2 & 0.27 & 0.7 \\
5 & 5927.5, 7045.6 & 0.28 & 0.66 \\
7 & 5535.7, 5541.0, 5545.0 & 0.28 & 0.66 \\
8 & 6330.4, 7721.0 & 0.23 & 0.64 \\
1 & 7357.2, 7585.6 & 0.05 & 0.33 \\
11 & 7321.1, 7832.7 & 0.00 & 0.29 \\
6 & 7334.3, 7935.3 & 0.02 & 0.27 \\
9 & 7099.5, 7105.9 & 0.00 & 0.14 \\
12 & 6998.7, 7249.3 & 0.00 & 0.14 \\
10 & 6837.7, 7562.2 & 0.00 & 0.12 \\

\hline 
\hspace{-0.1cm}
\tablenotetext{a}{Probability for perfect recovery.}
\tablenotetext{b}{Probability for TRUE POSITIVE = P(0) + P(1)}
\end{tabular} 
\label{t:3sig_groups_dibs}
\end{table}

As found for atomic and molecular lines in the previous section, the probability of a group to be perfectly recovered is essentially null. We find that 6 out of the 12 groups of DIBs have a probability that is higher than 0.5 to be a TP group, similar to the numbers obtained for the atomic and molecular groups. Following the simulations and the atomic and molecular clustering we suggest that groups 2, 3, 4, 5, 7 and 8 in table \ref{t:3sig_groups_dibs} are spectroscopic families. 
We use equation \ref{eq:8} to extract the real correlations out of the normalized correlations, and find that the DIBs in groups 2,3,4,5,7 and 8 have pairwise partial correlations of $0.45-0.6$ while the DIBs in the remaining groups (the groups with probabilities to be TP groups that is smaller than 0.5) have pairwise partial correlations of $0.3-0.4$.

One can see that none of the strongest, most extensively studied DIBs, is present in the final groups we present. We show in section \ref{s:dibs_corr_red} that the DIBs 5780.6\,\AA, 5797.1\,\AA, 6613.7\,\AA\, and 8621.1\,\AA\, have high correlations with $\mathrm{E}(B-V)$, it is therefore probable that most of their signal is removed in the process of computing the partial correlation. For the 6196.2\,\AA\, DIB a correlation with $\mathrm{E}(B-V)$ cannot be measured since its slope is not above our threshold. We also compute the correlation between the 6613.7\,\AA\, and 6196.2\,\AA\, DIB pair, which has been shown to correlate strongly. We find a negligible correlation of 0.0031 between the two DIBs which is expected since the DIB 6196.2\,\AA\, is undetected in our spectra, as discussed in paper1.

Some further clues may be gained by clustering DIBs together with the atomic and molecular lines. We cluster the 197 DIBs in our catalog and the 114 molecular and atomic absorption lines and obtain a 311 X 311 normalized partial correlation matrix. Our algorithm finds 27 groups of absorption lines, two of them are clearly false positives since they contain absorption lines that arise from different molecules, and their probabilities are 0.21 and 0.17, reflecting their dubiousness. 16 clusters have a TP probability that is higher than 0.5, and essentially recover the groups found by clustering them separately. The only new result are four clusters that include a mix of molecules ($\mathrm{C}_2$ or CN) and DIBs, and we list them in table \ref{t:dibs_mols_final}. 

\subsection{Optimization: more clusters, less confidence}\label{s:opt}

Our ultimate goal was (and still is) to cluster all the DIBs into families. However, we are limited by our SNR and systematic effects that are beyond our control. Our simulations indicate, based on the slope and the correlation with dust extinction, that we should be able in principle to cluster 76 molecular and atomic absorption lines, and 57 DIBs with reasonable probabilities. The fact that we only see 22 atomic and molecular lines and 26 DIBs clustering is an indication that their correlation is too weak for us to measure. 

A solution that would increase the number (and size) of clusters, would be to lower the threshold for grouping, currently set at $3\sigma$. Using a threshold of $2\sigma$ we obtain 62 DIBs, and 54 atomic and molecular absorption lines that are clustered into groups, which are the majority of the detected lines. However, the price one pays is a higher false positive rate and lower confidence for the clusters obtained. Future, higher SNR, data may allow for more robust clustering of more lines. For reference, we include a table of the $2\sigma$ DIBs clusters (Table \ref{t:2sig_groups_dibs}), which may be indicative, if somewhat spurious. 

\begin{table}
\setlength{\tabcolsep}{3pt}
\caption{DIBs groups for $2\sigma$ threshold} 

\begin{tabular}{lccc} 
\hline \hline 

Group ID & DIBs[\AA] & P(0)\tablenotemark{a} & P(TP)\tablenotemark{a} \\ [0.5ex]

\hline

4 & 6497.5, 6862.5, 7569.7 & 0.28 & 0.67 \\
6 & 5900.6, 5923.4, 6317.6, 6325.1 & 0.31 & 0.66 \\
8 & 5927.5, 7045.6 & 0.29 & 0.60 \\
11 & 5535.7, 5541,  5545 & 0.29 & 0.60 \\
12 & 6330.4, 7721 & 0.25 & 0.59 \\
22 & 5524.9, 6845.3 & 0.16 & 0.54 \\
1 & 6834.5, 7401.7 & 0.12 & 0.49 \\
5 & 6362.4, 6827.3 & 0.12 & 0.49 \\
7 & 7276.7, 7558.2, 8026.2 & 0.12 & 0.49 \\
26 & 6236.6, 6741. & 0.0 & 0.31 \\
3 & 7357.2, 7585.6 & 0.036 & 0.28 \\
20 & 7915.1, 8763.5 & 0.0 & 0.28 \\
9 & 7153.8, 7334.3, 7935.3 & 0.0 & 0.27 \\
19 & 6993.2, 6998.7, 7249.3 & 0.0 & 0.27 \\
15 & 6270.1, 8038.5 & 0.0 & 0.26 \\
27 & 6536.4, 7695.9 & 0.0 & 0.26 \\
2 & 6978.5, 7330.2 & 0.0 & 0.24 \\
25 & 4880.4, 6460.3 & 0.016 & 0.21 \\
14 & 5418.4, 6944.5 & 0.0 & 0.20 \\
17 & 7321.1, 7832.7 & 0.0 & 0.20 \\
18 & 5719.4, 7705.9 & 0.0 & 0.20 \\
16 & 6837.7, 7562.2 & 0.0 & 0.15 \\
10 & 6860,  7366.6 & 0.0 & 0.074 \\
24 & 6689.3, 6694.5 & 0.0 & 0.074 \\
13 & 5947.3, 7099.5, 7105.9 & 0.0 & 0.071 \\
21 & 5362.1, 7349.8 & 0.0 & 0.019 \\
23 & 4501.8, 6632.9 & 0.0 & 0.019 \\

\hline 
\hspace{-0.1cm}
\tablenotetext{a}{Probability for perfect recovery.}
\tablenotetext{b}{Probability for TRUE POSITIVE = P(0) + P(1)}
\end{tabular} 
\label{t:2sig_groups_dibs}
\end{table}

\begin{table}
\setlength{\tabcolsep}{3pt}
\caption{DIBs spectroscopic families and their associated molecules\tablenotemark{*}} 

\begin{tabular}{lccc} 
\hline \hline 

DIBs [\AA] & P(TP)\tablenotemark{a} & Molecule & P(TP)\tablenotemark{b} \\ [0.5ex]

\hline

6317.6, 6325.1 & 0.74 & C2 & 0.74 \\
6497.5, 6862.5, 7569.7 & 0.7 & C2 & 0.65 \\
7276.7, 8026.2 & 0.7 & C2 & 0.69 \\
5927.5, 7045.6 & 0.66 & CN & 0.57 \\

\hline 
\hspace{-0.1cm}
\tablenotetext{*}{The table contains only groups with $P(TP) > 0.5$ and that are associated with a molecule with $P(TP) > 0.5$.}
\tablenotetext{a}{Probability for the DIB group to be true positive.}
\tablenotetext{b}{Probability for the group to be associated with the molecule.}
\end{tabular} 
\label{t:dibs_mols_final}
\end{table}

\section{Conclusions and Discussion}\label{s:conc}

In this paper we systematically study the DIBs at high Galactic latitude, with a sampling and coverage never remotely approached before. Our findings can be separated into two. Based on the correlation of the DIBs with dust maps we find the following. 

\begin{itemize}

	\item We measure the correlation of 142 DIBs with dust extinction and find that only 19 of them show a significant correlation that is higher than 0.7. Some were measured in previous studies ($5705$\,\AA, $5780$\,\AA, $5797.1$\,\AA, $6284.3$\,\AA, $6613.7$\,\AA\, and $8621.1$\,\AA). Most of the DIBs do not correlate with dust in the $\mathrm{E}(B-V)$ range 0 -- 0.18 Mag.

	\item We find a pair of DIBs, $5925.9$\,\AA\, and $5927.5$\,\AA, with significant negative correlation of their depth with $\mathrm{E}(B-V)$. The carrier of these lines is probably depleted in dusty clouds.

	\item We find that the DIBs $4428$\,\AA, $5512$\,\AA, $5850$\,\AA, and $6445$\,\AA\, are absent in our data and the DIBs $4762$\,\AA, $4964$\,\AA, $5487$\,\AA, $6090$\,\AA\ and $6379$\,\AA\, correlate weakly with dust. This group of absorption lines traverses previously established grouping based on dust and UV radiation. Moreover, previous studies reported high correlations with dust for these lines. Since we do detect and measure strong correlations for DIBs with similar strengths, we suggest that their non detection (or low correlation) is connected to the environment we probe at high Galactic latitude.

\end{itemize}

We subsequently apply a Hierarchical Clustering algorithm to the matrices of  pairwise correlation between lines. 

\begin{itemize}
	
	\item Testing our algorithm on 114 known atomic and molecular lines, we find that it is able to cluster most of the detected atomic and molecular absorption lines in our data. We obtain 11 groups of atomic and molecular lines ($\mathrm{C}_2$, CN and NaID), 9 are robust. Furthermore, we find that the only group that contains absorption lines that originate from different molecules is also the group with the lowest probability to be correct.
	
	\item Our simulations show that the algorithm we develop is mostly sensitive to DIBs with low correlation with $\mathrm{E}(B-V)$. This allows us to study weak DIBs which are typically barely studied and divide them into spectroscopic families.
	
	\item We find 6 robust groups of DIBs (table \ref{t:3sig_groups_dibs}). We suggest that each of these groups is a spectroscopic family formed by a different carrier.
	
	\item By clustering DIBs with known atomic and molecular lines we find that 4 of the 6 DIB groups are associated with molecular absorption lines. The groups (1) 6317.6\,\AA\, and 6325.1\,\AA\, (2) 6497.5\,\AA, 6862.5\,\AA\, and 7569.7\,\AA\, (3) 7276.7\,\AA\, and 8026.2\,\AA\, are associated with $\mathrm{C}_2$ absorption lines, and the group (4) 5927.5\,\AA\, and 7045.6\,\AA\, is associated with CN absorption lines. 

\end{itemize}

As the study of DIBs evolves to systematic observations of thousands of lines of sight, the methodology needs to progress as well. With the added complexity also comes the promise of greater discoveries. In this study we clustered DIBs based on their partial correlation, after removing their individual correlation with dust extinction. With larger datasets (e.g., from the Dark Energy Survey Instrument \citealt{levi13}), and/or better SNR per spectrum \citep{apellaniz14}, one could compare DIBs strength to additional species. One could measure the correlations of DIBs with H\,I and H$_2$ (via CO), PAHs, and also Galactic latitude, for example. These correlations can then again be fed to a clustering algorithm, and perhaps bring us a critical step closer to robust DIBs families. 

\section*{Acknowledgments}
We thank the anonymous referee for their review of this manuscript.
We further thank 
Anja C. Andersen,
A. Cohen,
N. Levy, 
and A. Loeb
for useful comments or valuable advice regarding different aspects of this work. 

D.P. acknowledges the support of the Alon fellowship for outstanding young researchers, and of the Raymond and Beverly Sackler Chair for young scientists. D. B. and D. P. thank the Dark Cosmology Center which is funded by the Danish National Research Foundation for hosting them while working on this topic.

The bulk of our computations was performed on the resources of the National Energy Research Scientific Computing Center, which is supported by the Office of Science of the U.S. Department of Energy under Contract No. DE-AC02-05CH11231, using the open source scientific database SciDB\footnote{www.scidb.org}. The  spectroscopic analysis was made using IPython \citep{perez07}. We also used these Python packages:  pyspeckit\footnote{www.pyspeckit.bitbucket.org}, healpy\footnote{www.healpy.readthedocs.org}, and astropy\footnote{www.astropy.org/}.

This work made extensive use of SDSS-III\footnote{www.sdss3.org} data. Funding for SDSS-III has been provided by the Alfred P. Sloan Foundation, the Participating Institutions, the National Science Foundation, and the U.S. Department of Energy Office of Science. SDSS-III is managed by the Astrophysical Research Consortium for the Participating Institutions of the SDSS-III Collaboration including the University of Arizona, the Brazilian Participation Group, Brookhaven National Laboratory, Carnegie Mellon University, University of Florida, the French Participation Group, the German Participation Group, Harvard University, the Instituto de Astrofisica de Canarias, the Michigan State/Notre Dame/JINA Participation Group, Johns Hopkins University, Lawrence Berkeley National Laboratory, Max Planck Institute for Astrophysics, Max Planck Institute for Extraterrestrial Physics, New Mexico State University, New York University, Ohio State University, Pennsylvania State University, University of Portsmouth, Princeton University, the Spanish Participation Group, University of Tokyo, University of Utah, Vanderbilt University, University of Virginia, University of Washington, and Yale University.


\bibliographystyle{mn2e}
\bibliography{ref}

\begin{thebibliography}{60}
\expandafter\ifx\csname natexlab\endcsname\relax\def\natexlab#1{#1}\fi

\bibitem[{{Baron} {et~al}\mbox{.}(2015){Baron}, {Poznanski}, {Watson}, {Yao},
  \& {Prochaska}}]{baron15}
{Baron} D., {Poznanski} D., {Watson} D., {Yao} Y., {Prochaska} J.~X., 2015,
  \mnras, 447, 545

\bibitem[{{Bryndal} \& {Wszo{\l}ek}(2007)}]{bryndal07}
{Bryndal} K., {Wszo{\l}ek} B., 2007, in 14th Young Scientists Conference on
  Astronomy and Space Physics, {Ivashchenko} G., {Golovin} A., eds., pp. 22--24

\bibitem[{{Cami} {et~al}\mbox{.}(1997){Cami}, {Sonnentrucker}, {Ehrenfreund},
  \& {Foing}}]{cami97}
{Cami} J., {Sonnentrucker} P., {Ehrenfreund} P., {Foing} B.~H., 1997, \aap,
  326, 822

\bibitem[{{Chlewicki} {et~al}\mbox{.}(1986){Chlewicki}, {van der Zwet}, {van
  Ijzendoorn}, {Greenberg}, \& {Alvarez}}]{chlewicki86}
{Chlewicki} G., {van der Zwet} G.~P., {van Ijzendoorn} L.~J., {Greenberg}
  J.~M., {Alvarez} P.~P., 1986, \apj, 305, 455

\bibitem[{{Cordiner} {et~al}\mbox{.}(2011){Cordiner}, {Cox}, {Evans},
  {Trundle}, {Smith}, {Sarre}, \& {Gordon}}]{Cordiner11}
{Cordiner} M.~A., {Cox} N.~L.~J., {Evans} C.~J., {Trundle} C., {Smith} K.~T.,
  {Sarre} P.~J., {Gordon} K.~D., 2011, \apj, 726, 39

\bibitem[{{Cordiner} {et~al}\mbox{.}(2008){Cordiner}, {Cox}, {Trundle},
  {Evans}, {Hunter}, {Przybilla}, {Bresolin}, \& {Salama}}]{cordiner08}
{Cordiner} M.~A., {Cox} N.~L.~J., {Trundle} C., {Evans} C.~J., {Hunter} I.,
  {Przybilla} N., {Bresolin} F., {Salama} F., 2008, \aap, 480, L13

\bibitem[{{Cox} {et~al}\mbox{.}(2006){Cox}, {Cordiner}, {Cami}, {Foing},
  {Sarre}, {Kaper}, \& {Ehrenfreund}}]{cox06b}
{Cox} N.~L.~J., {Cordiner} M.~A., {Cami} J., {Foing} B.~H., {Sarre} P.~J.,
  {Kaper} L., {Ehrenfreund} P., 2006, \aap, 447, 991

\bibitem[{{Cox} {et~al}\mbox{.}(2005){Cox}, {Kaper}, {Foing}, \&
  {Ehrenfreund}}]{cox05}
{Cox} N.~L.~J., {Kaper} L., {Foing} B.~H., {Ehrenfreund} P., 2005, \aap, 438,
  187

\bibitem[{{Cox} \& {Patat}(2008)}]{cox08}
{Cox} N.~L.~J., {Patat} F., 2008, \aap, 485, L9

\bibitem[{{Ehrenfreund} {et~al}\mbox{.}(2002){Ehrenfreund}, {Cami},
  {Jim{\'e}nez-Vicente}, {Foing}, {Kaper}, {van der Meer}, {Cox},
  {D'Hendecourt}, {Maier}, {Salama}, {Sarre}, {Snow}, \&
  {Sonnentrucker}}]{ehrenfreund02}
{Ehrenfreund} P. {et~al.}, 2002, \apjl, 576, L117

\bibitem[{{Friedman} {et~al}\mbox{.}(2011){Friedman}, {York}, {McCall},
  {Dahlstrom}, {Sonnentrucker}, {Welty}, {Drosback}, {Hobbs}, {Rachford}, \&
  {Snow}}]{friedman11}
{Friedman} S.~D. {et~al.}, 2011, \apj, 727, 33

\bibitem[{{Fulara} \& {Krelowski}(2000)}]{fulara00}
{Fulara} J., {Krelowski} J., 2000, New Astronomy Reviews, 44, 581

\bibitem[{{Galazutdinov} {et~al}\mbox{.}(2002){Galazutdinov}, {Moutou},
  {Musaev}, \& {Kre{\l}owski}}]{galazutdinov02a}
{Galazutdinov} G., {Moutou} C., {Musaev} F., {Kre{\l}owski} J., 2002, \aap,
  384, 215

\bibitem[{{Heger}(1922)}]{heger22}
{Heger} M.~L., 1922, Lick Observatory Bulletin, 10, 141

\bibitem[{{Herbig}(1995)}]{herbig95}
{Herbig} G.~H., 1995, \araa, 33, 19

\bibitem[{{Herbig} \& {Leka}(1991)}]{herbig91}
{Herbig} G.~H., {Leka} K.~D., 1991, \apj, 382, 193

\bibitem[{{Hobbs} {et~al}\mbox{.}(2008){Hobbs}, {York}, {Snow}, {Oka},
  {Thorburn}, {Bishof}, {Friedman}, {McCall}, {Rachford}, {Sonnentrucker}, \&
  {Welty}}]{hobbs08}
{Hobbs} L.~M. {et~al.}, 2008, \apj, 680, 1256

\bibitem[{{Jenniskens} \& {Desert}(1993)}]{jenniskens93}
{Jenniskens} P., {Desert} F.~X., 1993, \aap, 274, 465

\bibitem[{{Jenniskens} \& {Desert}(1994)}]{jenniskens94}
{Jenniskens} P., {Desert} F.-X., 1994, \aaps, 106, 39

\bibitem[{{Jenniskens} {et~al}\mbox{.}(1996){Jenniskens}, {Porceddu},
  {Benvenuti}, \& {Desert}}]{jenniskens96}
{Jenniskens} P., {Porceddu} I., {Benvenuti} P., {Desert} F.-X., 1996, \aap,
  313, 649

\bibitem[{{Josafatsson} \& {Snow}(1987)}]{josafatsson87}
{Josafatsson} K., {Snow} T.~P., 1987, \apj, 319, 436

\bibitem[{{Ka{\'z}mierczak} {et~al}\mbox{.}(2009){Ka{\'z}mierczak},
  {Gnaci{\'n}ski}, {Schmidt}, {Galazutdinov}, {Bondar}, \&
  {Kre{\l}owski}}]{kazmierczak09}
{Ka{\'z}mierczak} M., {Gnaci{\'n}ski} P., {Schmidt} M.~R., {Galazutdinov} G.,
  {Bondar} A., {Kre{\l}owski} J., 2009, \aap, 498, 785

\bibitem[{{Kos} \& {Zwitter}(2013)}]{kos13a}
{Kos} J., {Zwitter} T., 2013, \apj, 774, 72

\bibitem[{{Kos} {et~al}\mbox{.}(2013){Kos}, {Zwitter}, {Grebel}, {Bienayme},
  {Binney}, {Bland-Hawthorn}, {Freeman}, {Gibson}, {Gilmore}, {Kordopatis},
  {Navarro}, {Parker}, {Reid}, {Seabroke}, {Siebert}, {Siviero}, {Steinmetz},
  {Watson}, \& {Wyse}}]{kos13}
{Kos} J. {et~al.}, 2013, \apj, 778, 86

\bibitem[{{Kos} {et~al}\mbox{.}(2014){Kos}, {Zwitter}, {Wyse}, {Bienaym{\'e}},
  {Binney}, {Bland-Hawthorn}, {Freeman}, {Gibson}, {Gilmore}, {Grebel},
  {Helmi}, {Kordopatis}, {Munari}, {Navarro}, {Parker}, {Reid}, {Seabroke},
  {Sharma}, {Siebert}, {Siviero}, {Steinmetz}, {Watson}, \& {Williams}}]{kos14}
{Kos} J. {et~al.}, 2014, Science, 345, 791

\bibitem[{{Kre{\l}owski} {et~al}\mbox{.}(2010){Kre{\l}owski}, {Beletsky},
  {Galazutdinov}, {Ko{\l}os}, {Gronowski}, \& {LoCurto}}]{Kreowski10}
{Kre{\l}owski} J., {Beletsky} Y., {Galazutdinov} G.~A., {Ko{\l}os} R.,
  {Gronowski} M., {LoCurto} G., 2010, \apjl, 714, L64

\bibitem[{{Krelowski}, {Sneden} \& {Hittgen}(1995){Krelowski}, {Sneden}, \&
  {Hittgen}}]{krewski95}
{Krelowski} J., {Sneden} C., {Hittgen} D., 1995, \planss, 43, 1195

\bibitem[{{Krelowski} \& {Walker}(1987)}]{krelowski87}
{Krelowski} J., {Walker} G.~A.~H., 1987, \apj, 312, 860

\bibitem[{{Krelowski} \& {Westerlund}(1988)}]{krelowski88}
{Krelowski} J., {Westerlund} B.~E., 1988, \aap, 190, 339

\bibitem[{{Kroto} {et~al}\mbox{.}(1985){Kroto}, {Heath}, {Obrien}, {Curl}, \&
  {Smalley}}]{kroto85}
{Kroto} H.~W., {Heath} J.~R., {Obrien} S.~C., {Curl} R.~F., {Smalley} R.~E.,
  1985, \nat, 318, 162

\bibitem[{{Lan}, {M{\'e}nard} \& {Zhu}(2014){Lan}, {M{\'e}nard}, \&
  {Zhu}}]{lan14a}
{Lan} T.-W., {M{\'e}nard} B., {Zhu} G., 2014, ArXiv e-prints:1406.7284

\bibitem[{{Leidlmair} {et~al}\mbox{.}(2011){Leidlmair}, {Bartl}, {Sch{\"o}bel},
  {Denifl}, {Probst}, {Scheier}, \& {Echt}}]{leidlmair11}
{Leidlmair} C., {Bartl} P., {Sch{\"o}bel} H., {Denifl} S., {Probst} M.,
  {Scheier} P., {Echt} O., 2011, \apjl, 738, L4

\bibitem[{{Levi} {et~al}\mbox{.}(2013){Levi}, {Bebek}, {Beers}, {Blum}, {Cahn},
  {Eisenstein}, {Flaugher}, {Honscheid}, {Kron}, {Lahav}, {McDonald}, {Roe},
  {Schlegel}, \& {representing the DESI collaboration}}]{levi13}
{Levi} M. {et~al.}, 2013, ArXiv e-prints:1308.0847

\bibitem[{{Maier} {et~al}\mbox{.}(2011){Maier}, {Walker}, {Bohlender},
  {Mazzotti}, {Raghunandan}, {Fulara}, {Garkusha}, \& {Nagy}}]{Maier11}
{Maier} J.~P., {Walker} G.~A.~H., {Bohlender} D.~A., {Mazzotti} F.~J.,
  {Raghunandan} R., {Fulara} J., {Garkusha} I., {Nagy} A., 2011, \apj, 726, 41

\bibitem[{{Ma{\'{\i}}z Apell{\'a}niz} {et~al}\mbox{.}(2014){Ma{\'{\i}}z
  Apell{\'a}niz}, {Sota}, {Barb{\'a}}, {Morrell}, {Pellerin}, {Alfaro}, \&
  {Sim{\'o}n-D{\'{\i}}az}}]{apellaniz14}
{Ma{\'{\i}}z Apell{\'a}niz} J., {Sota} A., {Barb{\'a}} R.~H., {Morrell} N.~I.,
  {Pellerin} A., {Alfaro} E.~J., {Sim{\'o}n-D{\'{\i}}az} S., 2014, in IAU
  Symposium, Vol. 297, IAU Symposium, {Cami} J., {Cox} N.~L.~J., eds., pp.
  117--120

\bibitem[{{McCall}, {Drosback} \& {Thorburn}(2010){McCall}, {Drosback}, \&
  {Thorburn}}]{mccall10}
{McCall} B.~J., {Drosback} M.~M., {Thorburn} J.~A., 2010, \apj, 708, 1628

\bibitem[{{McIntosh} \& {Webster}(1993)}]{mcintosh93}
{McIntosh} A., {Webster} A., 1993, \mnras, 261, L13

\bibitem[{{Merrill}(1936)}]{merrill36}
{Merrill} P.~W., 1936, Contributions from the Mount Wilson Observatory /
  Carnegie Institution of Washington, 536, 1

\bibitem[{{Merrill} \& {Wilson}(1938)}]{merrill38}
{Merrill} P.~W., {Wilson} O.~C., 1938, \apj, 87, 9

\bibitem[{{Motylewski} {et~al}\mbox{.}(2000){Motylewski}, {Linnartz},
  {Vaizert}, {Maier}, {Galazutdinov}, {Musaev}, {Kre{\l}owski}, {Walker}, \&
  {Bohlender}}]{Motylewski00}
{Motylewski} T. {et~al.}, 2000, \apj, 531, 312

\bibitem[{{Moutou} {et~al}\mbox{.}(1999){Moutou}, {Kre{\l}owski},
  {D'Hendecourt}, \& {Jamroszczak}}]{moutou99}
{Moutou} C., {Kre{\l}owski} J., {D'Hendecourt} L., {Jamroszczak} J., 1999,
  \aap, 351, 680

\bibitem[{P\'erez \& Granger(2007)}]{perez07}
P\'erez F., Granger B.~E., 2007, Computing in Science and Engineering, 9, 21

\bibitem[{{Porceddu}, {Benvenuti} \& {Krelowski}(1991){Porceddu}, {Benvenuti},
  \& {Krelowski}}]{porceddu91}
{Porceddu} I., {Benvenuti} P., {Krelowski} J., 1991, \aap, 248, 188

\bibitem[{{Poznanski}, {Prochaska} \& {Bloom}(2012){Poznanski}, {Prochaska}, \&
  {Bloom}}]{poznanski12}
{Poznanski} D., {Prochaska} J.~X., {Bloom} J.~S., 2012, \mnras, 426, 1465

\bibitem[{{Salama} {et~al}\mbox{.}(1999){Salama}, {Galazutdinov},
  {Kre{\l}owski}, {Allamandola}, \& {Musaev}}]{salama99}
{Salama} F., {Galazutdinov} G.~A., {Kre{\l}owski} J., {Allamandola} L.~J.,
  {Musaev} F.~A., 1999, \apj, 526, 265

\bibitem[{{Sarre}(2006)}]{sarre06}
{Sarre} P.~J., 2006, Journal of Molecular Spectroscopy, 238, 1

\bibitem[{Savitzky \& Golay(1964)}]{savitzky64}
Savitzky A., Golay M. J.~E., 1964, Analytical Chemistry, 36, 1627

\bibitem[{{Schlegel}, {Finkbeiner} \& {Davis}(1998){Schlegel}, {Finkbeiner}, \&
  {Davis}}]{schlegel98}
{Schlegel} D.~J., {Finkbeiner} D.~P., {Davis} M., 1998, \apj, 500, 525

\bibitem[{{van Loon} {et~al}\mbox{.}(2013){van Loon}, {Bailey}, {Tatton},
  {Ma{\'{\i}}z Apell{\'a}niz}, {Crowther}, {de Koter}, {Evans},
  {H{\'e}nault-Brunet}, {Howarth}, {Richter}, {Sana}, {Sim{\'o}n-D{\'{\i}}az},
  {Taylor}, \& {Walborn}}]{van-loon13}
{van Loon} J.~T. {et~al.}, 2013, \aap, 550, A108

\bibitem[{{van Loon} {et~al}\mbox{.}(2009){van Loon}, {Smith}, {McDonald},
  {Sarre}, {Fossey}, \& {Sharp}}]{vavloon09}
{van Loon} J.~T., {Smith} K.~T., {McDonald} I., {Sarre} P.~J., {Fossey} S.~J.,
  {Sharp} R.~G., 2009, \mnras, 399, 195

\bibitem[{{Vos} {et~al}\mbox{.}(2011){Vos}, {Cox}, {Kaper}, {Spaans}, \&
  {Ehrenfreund}}]{vos11}
{Vos} D.~A.~I., {Cox} N.~L.~J., {Kaper} L., {Spaans} M., {Ehrenfreund} P.,
  2011, \aap, 533, A129

\bibitem[{{Welty} {et~al}\mbox{.}(2006){Welty}, {Federman}, {Gredel},
  {Thorburn}, \& {Lambert}}]{welty06}
{Welty} D.~E., {Federman} S.~R., {Gredel} R., {Thorburn} J.~A., {Lambert}
  D.~L., 2006, \apjs, 165, 138

\bibitem[{{Weselak} {et~al}\mbox{.}(2001){Weselak}, {Fulara}, {Schmidt}, \&
  {Kre{\l}owski}}]{weselak01}
{Weselak} T., {Fulara} J., {Schmidt} M.~R., {Kre{\l}owski} J., 2001, \aap, 377,
  677

\bibitem[{{Weselak} {et~al}\mbox{.}(2004){Weselak}, {Galazutdinov}, {Musaev},
  \& {Kre{\l}owski}}]{weselak04}
{Weselak} T., {Galazutdinov} G.~A., {Musaev} F.~A., {Kre{\l}owski} J., 2004,
  \aap, 414, 949

\bibitem[{{Weselak} {et~al}\mbox{.}(2014){Weselak}, {Galazutdinov}, {Sergeev},
  {Godunova}, {Ko{\l}os}, \& {Kre{\l}owski}}]{weselak14}
{Weselak} T., {Galazutdinov} G.~A., {Sergeev} O., {Godunova} V., {Ko{\l}os} R.,
  {Kre{\l}owski} J., 2014, Acta Astronomica, 64, 371

\bibitem[{{Wszolek}(2006)}]{wszolek06}
{Wszolek} B., 2006, in 13th Young Scientists' Conference on Astronomy and Space
  Physics, {Golovin} A., {Ivashchenko} G., {Simon} A., eds., p.~5

\bibitem[{{Wszo{\l}ek} \& {God{\l}owski}(2003)}]{wszoek03}
{Wszo{\l}ek} B., {God{\l}owski} W., 2003, \mnras, 338, 990

\bibitem[{{Xiang}, {Liu} \& {Yang}(2012){Xiang}, {Liu}, \& {Yang}}]{xiang12}
{Xiang} F., {Liu} Z., {Yang} X., 2012, \pasj, 64, 31

\bibitem[{{Zasowski} {et~al}\mbox{.}(2014){Zasowski}, {M{\'e}nard}, {Bizyaev},
  {Garc{\'{\i}}a P{\'e}rez}, {Johnson}, {Kinemuchi}, {Majewski}, {Shetrone}, \&
  {Wilson}}]{zasowski14}
{Zasowski} G. {et~al.}, 2014, ArXiv:1406.1195

\bibitem[{{Zhou} {et~al}\mbox{.}(2006){Zhou}, {Sfeir}, {Zhang}, {Hybertsen},
  {Steigerwald}, \& {Brus}}]{zhou06}
{Zhou} Z., {Sfeir} M.~Y., {Zhang} L., {Hybertsen} M.~S., {Steigerwald} M.,
  {Brus} L., 2006, \apjl, 638, L105

\end{thebibliography}


\appendix
\section{Flux - $\mathrm{E}(B-V)$ correlation simulations}\label{a:flux-red-sims}

Since our spectra undergo interpolation, normalization and stacking by $\mathrm{E}(B-V)$ it is likely that the initial correlation of the DIBs CD with $\mathrm{E}(B-V)$ will be affected. We wish to study the effects of the entire process on the correlation of various absorption lines and to find our detection limits. Our updated DIB catalog consists of 197 DIBs that spread over the SDSS wavelength range. We exclude DIBs with FWHM values that are less than $0.5$\,\AA\, using the average value that is reported in our updates DIB catalog and additional recent studies (e.g. \citealt{kos13a}). Apart from the blending that is caused by our low spectral resolution and our stacking of multiple sight-lines at various velocities, studies list DIBs that are partly to heavily blended. The latter arouse debates of whether these DIBs are separate and blended absorption lines that may be caused by different carriers or a complex structure that is generated by a single carrier. Blended lines must be examined and treated separately since their correlation is affected by both our stacking process and their neighbors correlation. We therefore divide our simulation to single absorption lines and blended absorption lines.

\subsection{single absorption lines}\label{s:sin_syn}

\begin{figure*}
\includegraphics[width=0.99\textwidth]{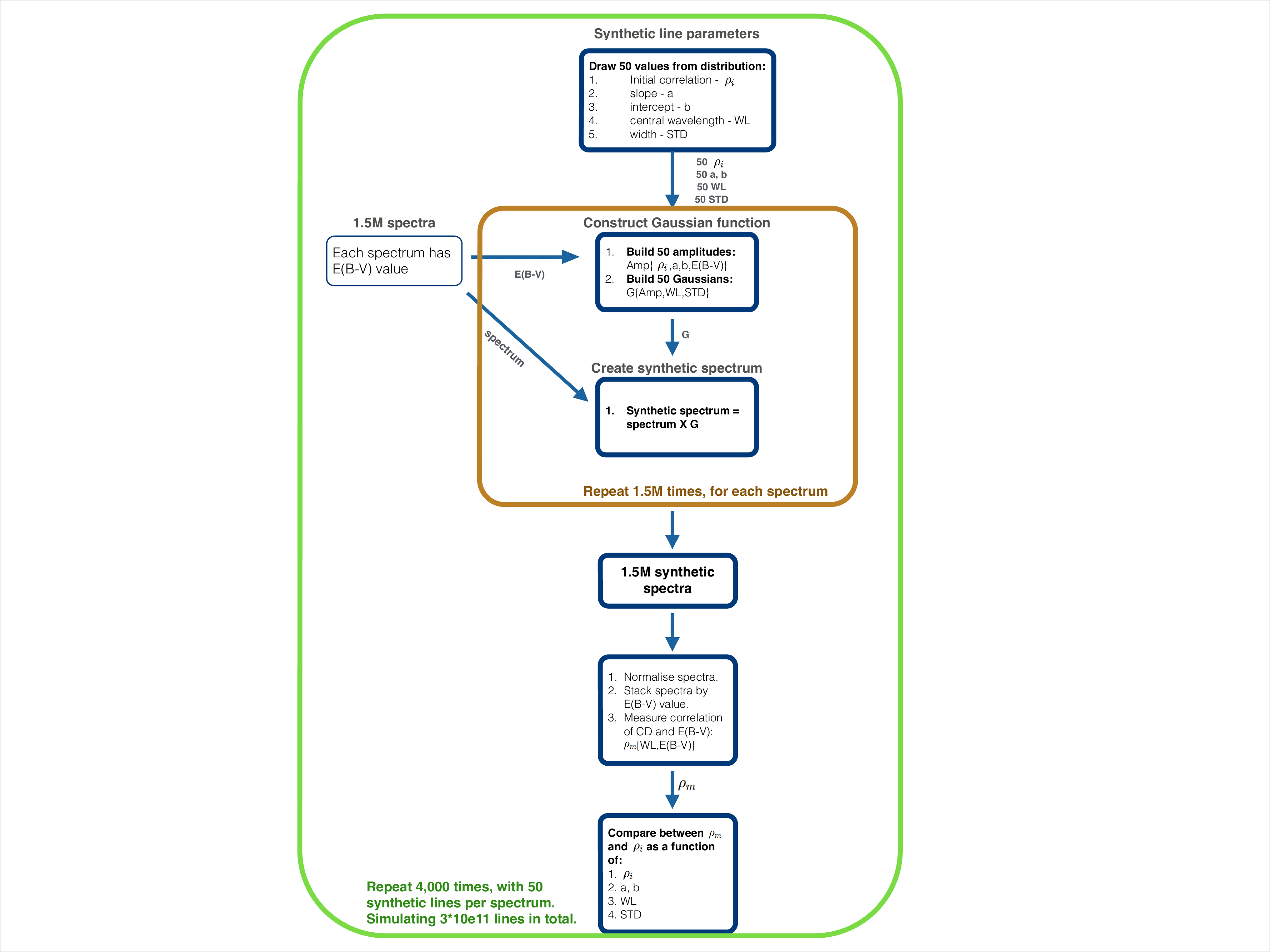}
\caption{Schematic representation of our basic simulation procedure.}\label{f:sims_a_sc}
\end{figure*}

Figure \ref{f:sims_a_sc} shows a schematic representation of our simulations, as described in details here. We construct synthetic lines that are represented by Gaussian profiles. To each line we assign a correlation with reddening, $\rho_{i}$, that is taken from a uniform distribution within the [$-1$, $1$] range. The correlation is constructed by varying the Gaussians' amplitude following the desired correlation, resulting in normally distributed amplitudes which can be both negative (absorption line) and positive (emission line) and is similar to the typical CD of DIBs (roughly 0.1\% - 3\% of the continuum level). We construct the amplitude as follows:

\begin{equation}\label{eq:6}
{A = [\rho_{i}\cdot\mathrm{E}(B-V) + \sqrt{1+\rho_{i}^{2}}\cdot\mathcal{N}(0,\sigma_{\mathrm{E}(B-V)})]\cdot a + b}
\end{equation}

Where $\mathcal{N}$ is Normal distribution with zero mean and standard deviation of $\sigma_{\mathrm{E}(B-V)}$, which is the standard deviation of our color excess distribution. The parameters \emph{a} and \emph{b} represent the slope and the intersect with the axis of the flux--$\mathrm{E}(B-V)$ relation and are sampled from a log-uniform distribution in the range [-7, 0]. We scale the Normal distribution to the color excess distribution in order to balance the correlation contribution and the noise contribution in equation \ref{eq:6}.

The standard deviation of the Gaussian is chosen from a uniform distribution in the range [1.06\,\AA, 2\,\AA] and is constant for different reddening values, the limits of the range are the 20th and 70th percentile respectively of the standard deviation that we measure for DIBs in paper1, we then convolve it with a 2.5\,\AA\, Gaussian to account for our resolution. This range does not represent broadest known DIBs (4428\,\AA\, for example) though we show in paper1 that these broad DIBs are not detected in our spectra. We set the central wavelength of each Gaussian from a uniform distribution in the range [3817\,\AA, 9206\,\AA] which covers all the SDSS wavelength grid in order to test for wavelength dependent biases and uncertainties.

We create synthetic spectra that contain the 50 Gaussians for every reddening value and multiply it by the 1.5M original spectra. The resulting spectra undergo the interpolation, normalization and stacking process discussed in section \ref{s:data}, after which we measure the correlation between the CD of the synthetic lines and reddening and compare it to the correlation we introduced. In order to construct a full image of the entire wavelength range we repeat the process of constructing Gaussians and injecting them to the original spectra 4000 times, in each repetition we assign new correlations, amplitudes, widths and central wavelengths, resulting in 200,000 synthetic lines that sample the entire parameter space well. The 200,000 synthetic lines are injected to the 1.5M raw spectra, which results in $3\cdot 10^{11}$ Gaussian injections in total.

\begin{figure}
\includegraphics[width=3.25in]{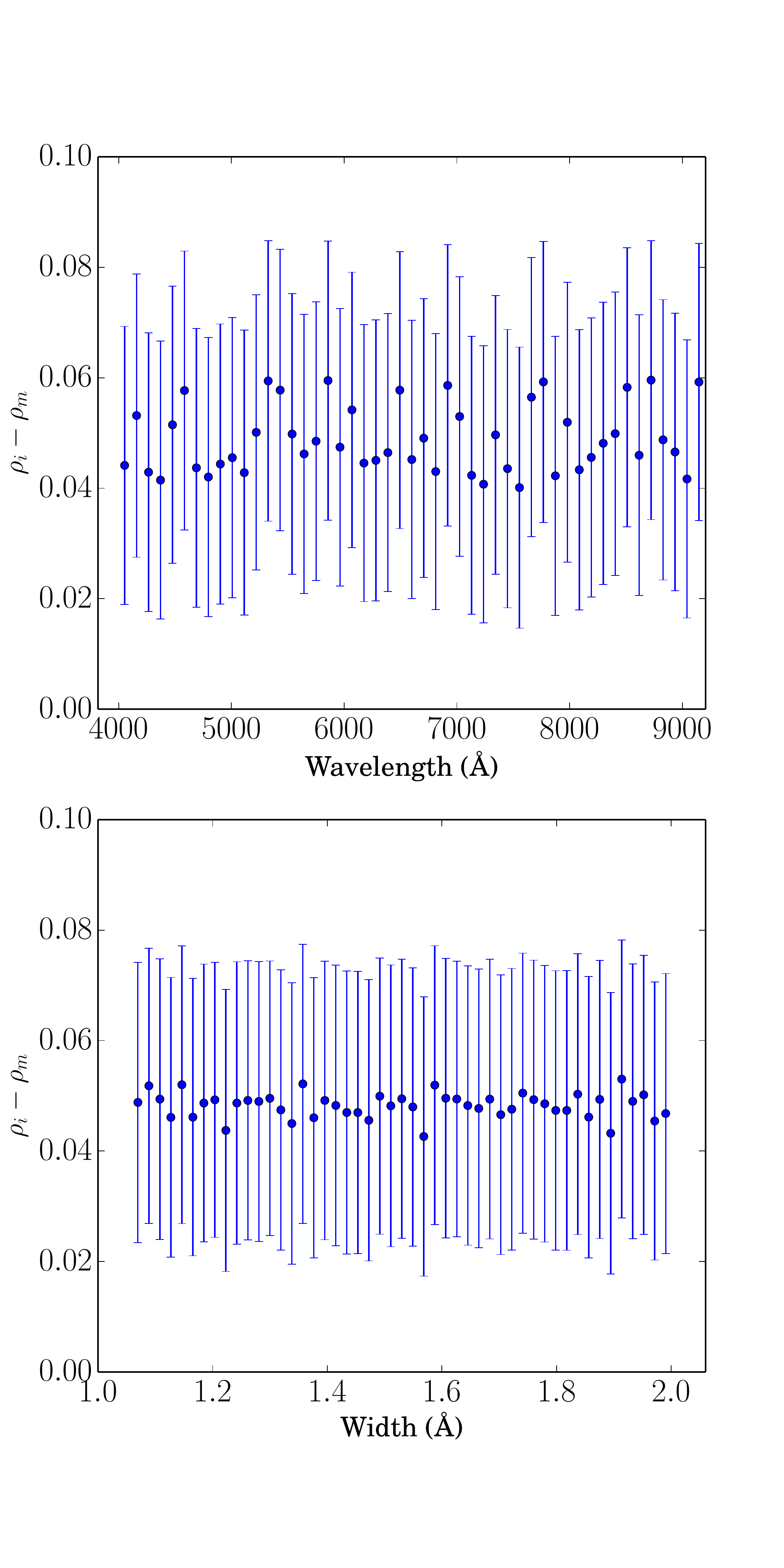}
\caption{Deviation of measured correlation from the introduced correlation as a function of the Gaussians central wavelength (top) and width (bottom).} \label{f:single_dep_wl_std}
\end{figure}

\begin{figure}
\includegraphics[width=3.25in]{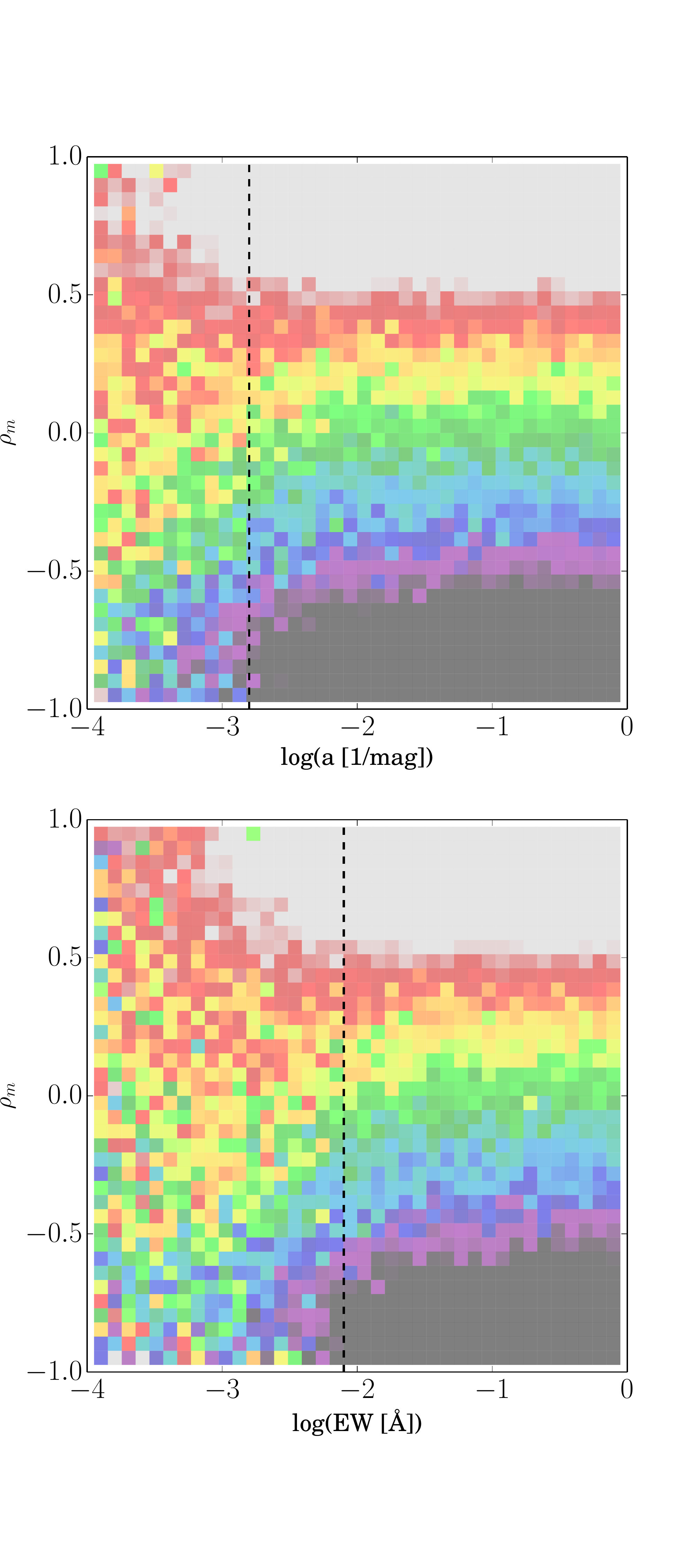}
\includegraphics[width=0.5\textwidth]{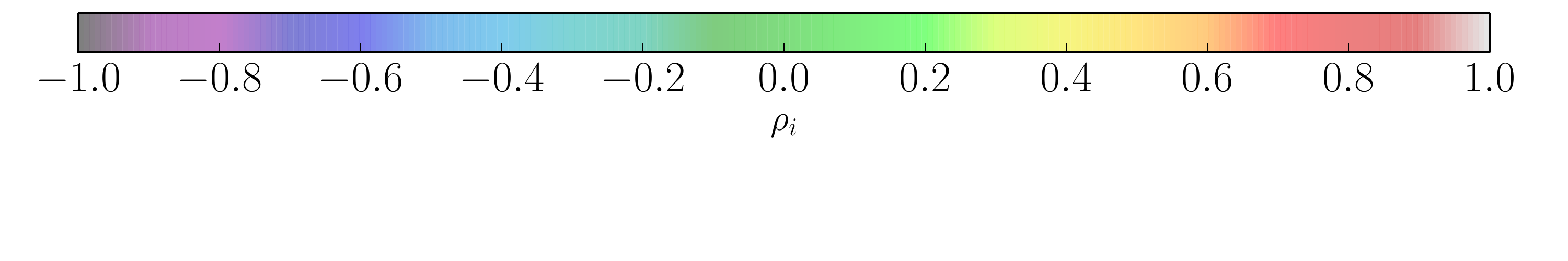}
\caption{Initial correlation as a function of measured correlation and slope (top) and EW (bottom) of the synthetic line. The thresholds of the correlation detections are 7 m\AA\, for the EW and $10^{-2.9} \frac{1}{mag} $ for the slope (black dashed).}\label{f:single_dep_a_ew}
\end{figure}

By comparing the measured correlation of the CD and reddening, which is measured after the injection and stacking process, to the initial correlation we construct, we find that the initial correlation can be recovered from the measured correlation regardless of the width of the Gaussian and its central wavelength as demonstrated in figure \ref{f:single_dep_wl_std}. Furthermore, we find that while the estimation uncertainty is independent of the intersect with the axis, b, for the range we use in equation \ref{eq:6}, it is sensitive to the initial slope, a. The estimation uncertainty is up to 10\% and roughly constant for slopes $a > 10^{-2.9} \frac{1}{mag}$. The top panel of figure \ref{f:single_dep_a_ew} presents the dependance of the correlation estimation on the slope. Therefore, the amplitude of the synthetic lines is dependent only on the slope (via equation \ref{eq:6}) and the mean EW of each synthetic line can be reconstructed from the mean amplitude and width of the Gaussian. The bottom panel of figure \ref{f:single_dep_a_ew} presents the correlation estimation as a function of the mean EW, where we use the mean amplitude of every line for the EW computation. Therefrom, the mean EW detection threshold is approximately 7 m\AA, which is similar to what we find in paper1. We chose to set $a = 10^{-2.9} \frac{1}{mag}$ as the detection threshold, above which the estimation uncertainty is no longer dependent on the slope.

\begin{figure}
\includegraphics[width=3.25in]{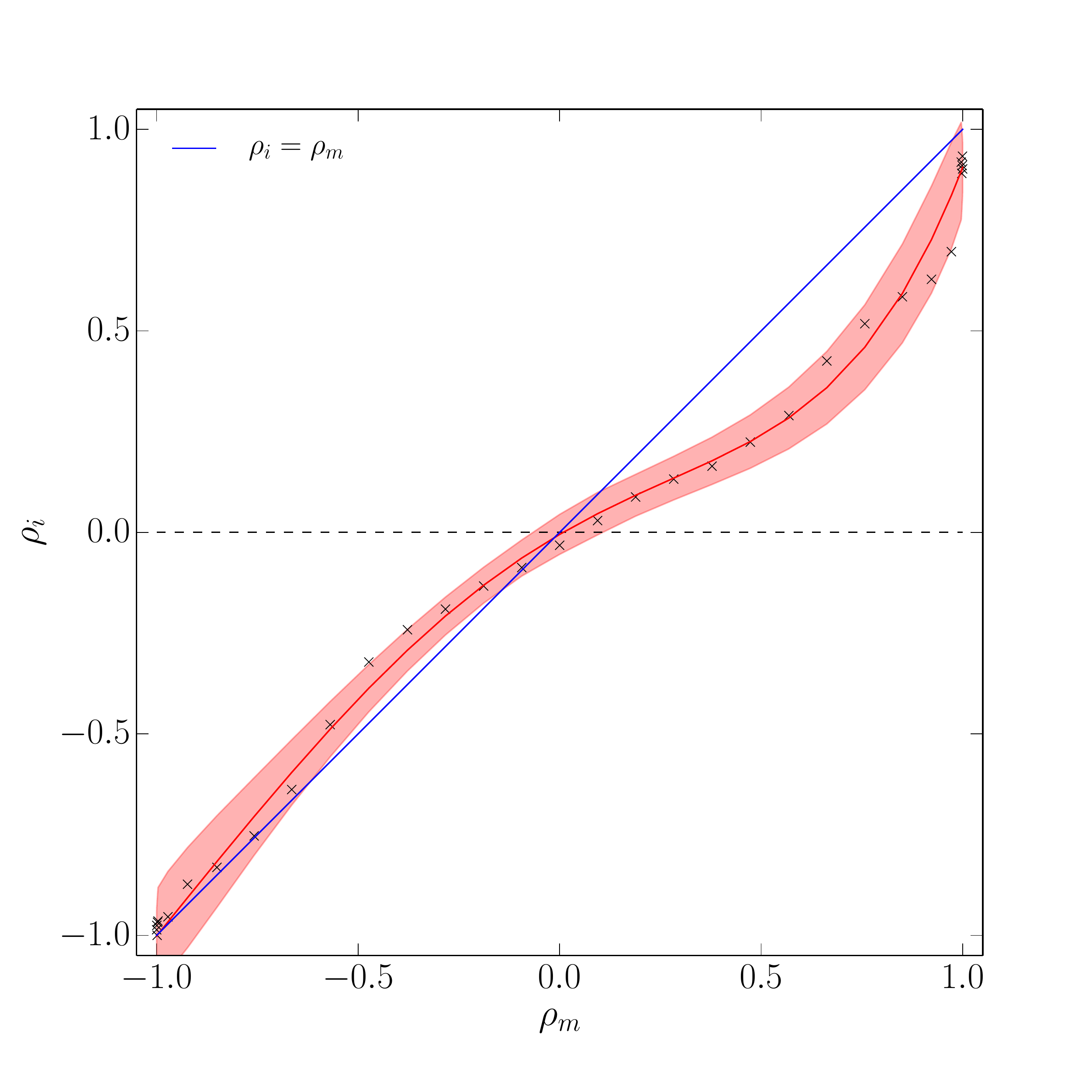}
\caption{Initial correlation as a function of measured correlation for isolated lines (black) and 5th polynomial fit that minimizes the reduced chi-square (red).}\label{f:single_sim}
\end{figure}

We divide the synthetic lines to 30 bins of measured correlation ($\rho_{m}$) and measure the median initial correlation ($\rho_{i}$) per bin, each bin contains 20,000--40,000 synthetic lines. For a measured correlation bin of zero the initial correlations are distributed normally around zero and as the measured correlation increases (decreases) the distribution of the initial correlation in each bin becomes left-skewed (right-skewed) since the correlation is blocked in the [-1,1] range. The uncertainty of each bin is defined as the MAD of the initial correlations in it. We present in figure \ref{f:single_sim} the initial correlation as a function of the measured correlation and a 5th polynomial fit which minimizes the reduced chi-square. This represents the effect of our entire process on the initial correlation of the flux and can be used to deduce the initial correlation of DIBs given their measured correlation. 
The dependance of the initial correlation on the measured correlation is not symmetric for positive and negative correlations. While negative initial correlations follow the measured correlation in an almost linear manner, positive initial correlations have a parabolic dependance of the measured correlation. This effect can be caused by residual negative correlation of the continuum-level with reddening -- absorption lines weaken with reddening (positive correlation) and the continuum-level decreases with reddening simultaneously which results in overestimation of positive correlations. On the other hand, negative correlations are overestimated when weak, where the correlation of the continuum-level with reddening is comparable to it, and recovers to the linear relation since the correlation of the lines becomes more significant as the correlation decreases to -1.

\subsection{blended absorption lines}\label{s:blend_syn}

We repeat the process done with the single lines but this time adding to each simulated line a companion line 0.5-3\,\AA\, away, with a uniform distance distribution. This distance distribution was constructed by trial and error, aiming to describe the blended lines entirely as we justify later in the section. Without prior knowledge we assume that both of the lines in a given pair can affect each others correlation with reddening and therefore we aim to execute a 2-dimensional binning (unlike the single synthetic lines for which we perform 1-dimensional binning) based on both of the neighbors correlations. In order to include a sufficient number of synthetic lines in each bin and result with a normally distributed measurement we repeat the process of injecting lines 40,000 times, a total of 2 million synthetic lines.

\begin{figure}
\includegraphics[width=3.25in]{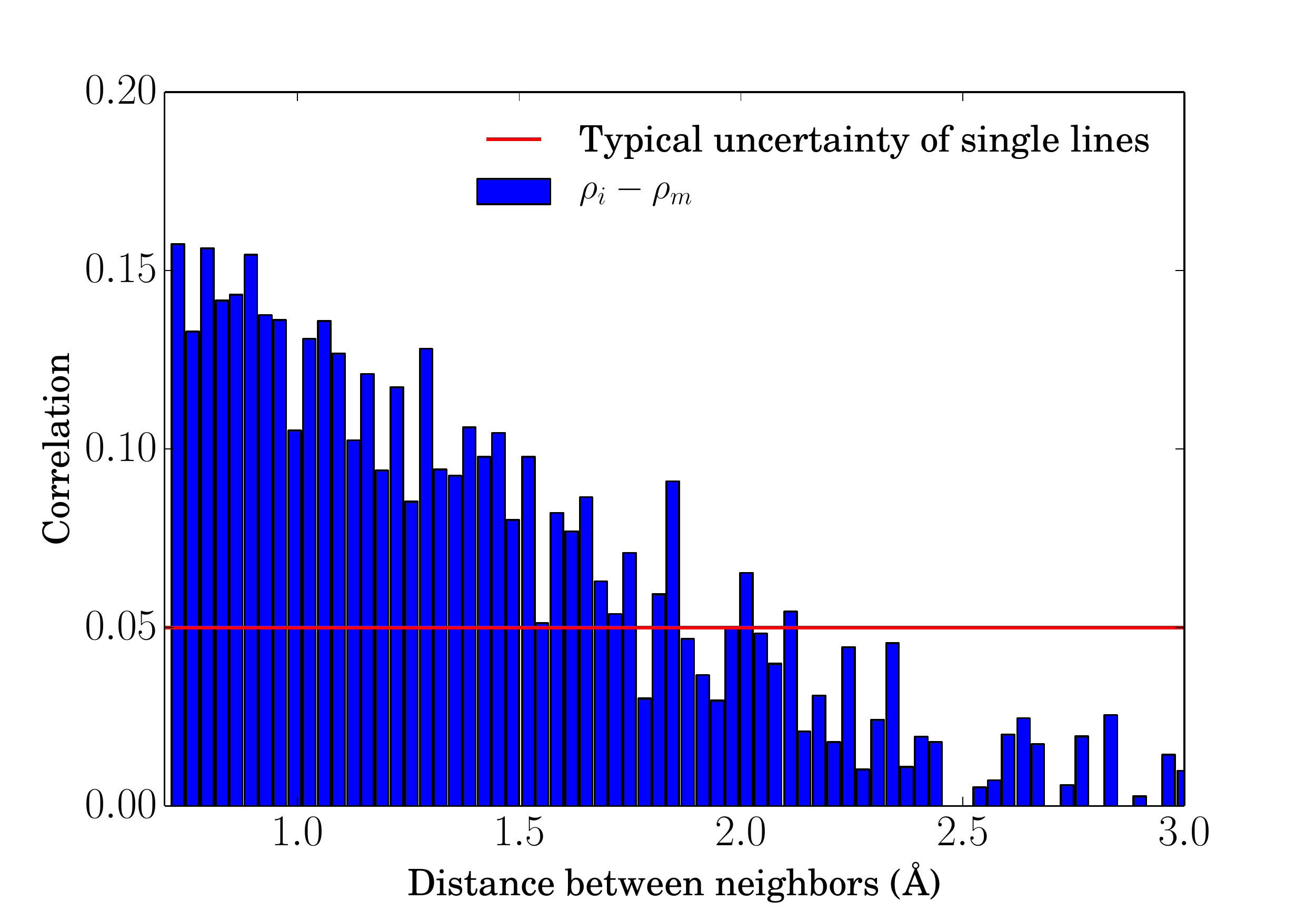}
\caption{Deviation in measured correlation ($\rho_{m}$) from initial correlation ($\rho_{i}$) as a function of distance between central wavelengths of the Gaussian pair (black) and the typical correlation estimation uncertainty of single lines (red).}\label{f:blend_dep}
\end{figure}

We compare between the initial correlations of the pair to the measured correlations as a function of the wavelength distance between the lines and find a deviation from the single line behavior for distances lower than 1.75\,\AA. As one can see in figure \ref{f:blend_dep}, pairs that are further apart than 1.75\,\AA\, exhibit a behavior that is similar to the single lines profile (figure \ref{f:single_sim}) with up to 5\% error, which is the typical uncertainty we measure for single lines. Pairs that are beneath the distance threshold show a mean deviation of up to 15\%.

In order to examine only blended lines we exclude all pairs for which the distance is higher than 1.75\,\AA, remaining with 1 million synthetic lines.
A comparison between the initial correlations with reddening of the pair to the measured correlation brings us to divide the pair population into two categories: pairs with the same correlation signs and pairs with opposite correlation signs. Pairs of initial correlations with the opposite sign tend to destroy each others' correlation almost entirely - for an absolute correlation range of [0, 0.9] the measured correlation degenerates to an almost constant value in the range [0, 0.05] while the absolute correlation range of [0.9, 1] results in measured correlation of [0.1, 0.4]. This degeneracy prevents us from deducing a function that converts measured correlation to initial correlation and we discard all the DIB pairs that fall into this category (5 pairs).
We find that pairs of DIBs with the same sign of initial correlation with reddening have a mutual effect on each other. The measured correlation with reddening of every line in the pair can be as high as the initial higher correlation but it can go down to zero regardless of the initial lower correlation (e.g, a pair with initial correlations of 0.9 and 0.7 can exhibit a measured correlation of 0.6 and 0.1). Furthermore, we find that the line with higher initial correlation presents the higher measured correlation on average. We therefore define the line with the higher correlation as Line1 and the line with the lower correlation as Line2 and examine how the initial correlation of Line1 and Line2 depends on the measured correlation of both of them. 

\begin{figure*}
\includegraphics[width=0.99\textwidth]{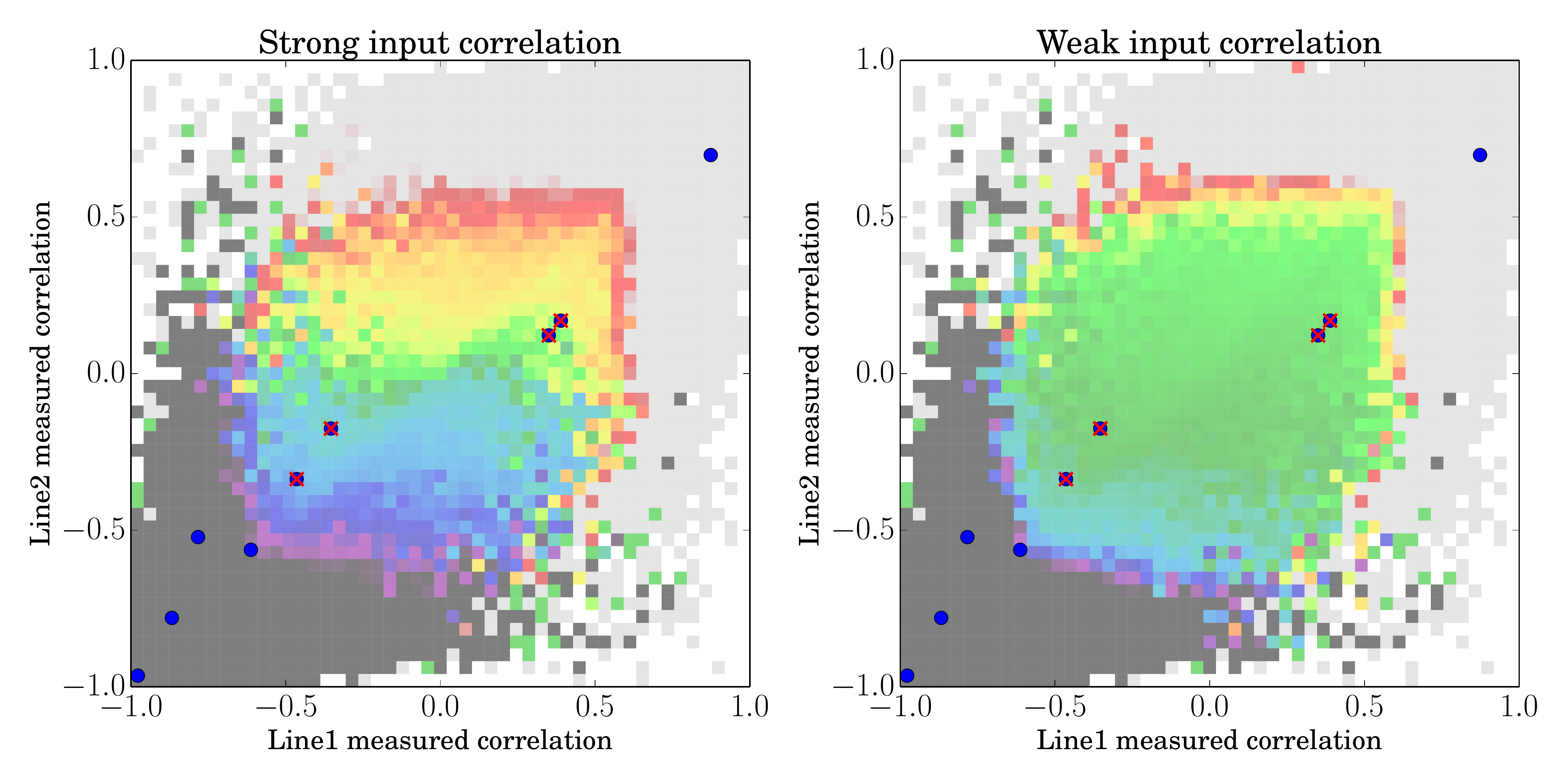}
\includegraphics[width=0.5\textwidth]{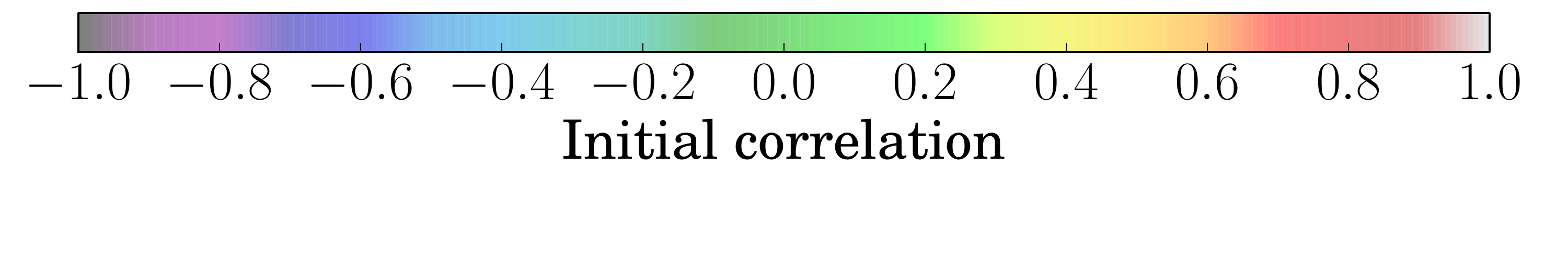}
\caption{Line1 (left) and Line2 (right) initial correlation as a function of measured correlations. The blue points are all the pairs of blended DIBs, from this we deduce the initial correlation of Line1 and Line2. Red crosses represent pairs of DIBs which we can not measure initial correlation for as the scatter is too big.}\label{f:blend_sim_vals}
\end{figure*}

\begin{figure*}
\includegraphics[width=0.99\textwidth]{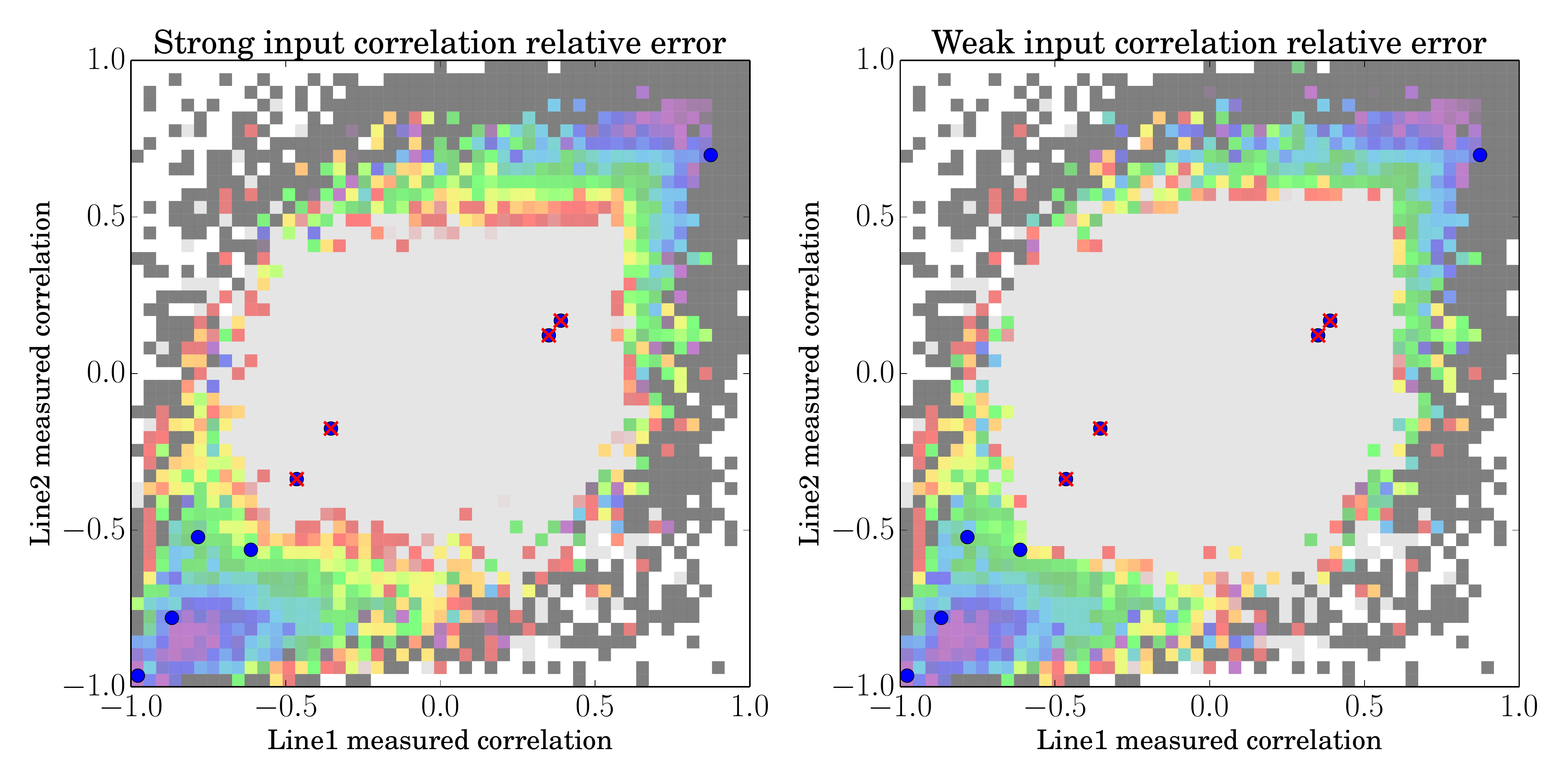}
\includegraphics[width=0.5\textwidth]{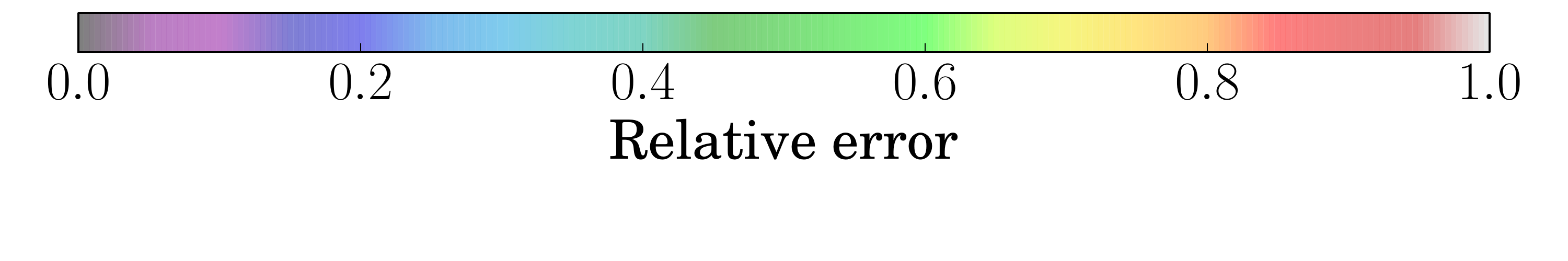}
\caption{Top panel: Line1 (left) and Line2 (right) relative uncertainty in every measured correlation bin. The blue points are all the pairs of blended DIBs, from this we deduce the initial correlation uncertainty of Line1 and Line2. Red crosses represent pairs of DIBs which we can not measure initial correlation for as the scatter is too big.}\label{f:blend_sim_errs}
\end{figure*}

\begin{figure}
\includegraphics[width=3.25in]{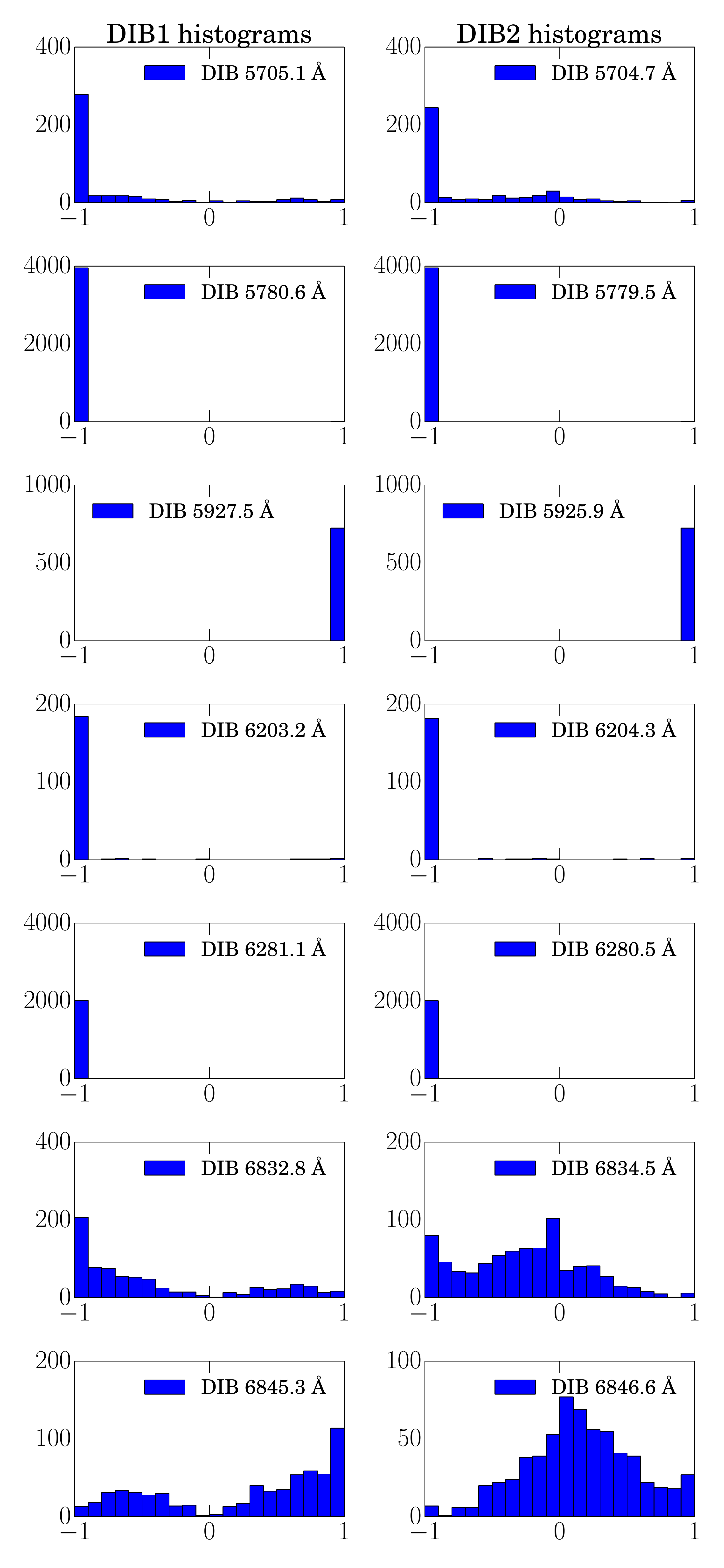}
\caption{Histograms of initial correlation of DIB1 (left panel) and DIB2 (right panel) for a 2D measured correlation bin. The 7 rows represent the 5 pairs of DIBs for which one can measure an initial correlation (represented by blue circles in figures \ref{f:blend_sim_vals} and \ref{f:blend_sim_errs}) and 2 pairs of DIBs which exhibit measured correlations that are lower than 0.5 (represented by blue circles with red crosses in figures \ref{f:blend_sim_vals} and \ref{f:blend_sim_errs}). One can see that the scatter for the latter is high as is indicated in figure \ref{f:blend_sim_errs}, their initial correlation uncertainty is therefore high (see table \ref{t:corr_red_strong}).}\label{f:blend_hist}
\end{figure}

We present in figures \ref{f:blend_sim_vals} and \ref{f:blend_sim_errs} the 2D binning of the synthetic lines. Figure \ref{f:blend_sim_vals} represents the initial correlation of Line1 (left) and Line2 (right) as a function of the measured correlations of the pair. Figure \ref{f:blend_sim_errs} represents the relative scatter per bin. The bins were constructed by the measured correlation values (50 bins for the range -1 to 1), thus the number of lines varies from bin to bin. As noted earlier, the bins that represent extreme correlations with opposite signs do not contain lines at all (white) regardless of the initial correlation. The 9 remaining DIB pairs are represented as blue circles on the figure. We find that beneath an absolute correlation of 0.5 the initial correlations cannot be recovered due to the degeneracy of the colormap and the high scatter per bin, we therefore further exclude 4 pairs that present such correlations, these are marked with red crosses in the figures. Furthermore, 2 of these pairs contain DIBs that are below our threshold for detection. The remaining 5 DIB pairs exhibit measured correlations that are higher than 0.5 and we therefore can recover their initial correlations by crossmatching their measured correlations on the plot. We chose the bins that match the 7 DIBs pairs that remain, 5 pairs with correlation that is higher than 0.5 and 2 pairs which we can not measure initial correlation for, and plot histograms of the initial correlation for them in figure \ref{f:blend_hist}. Every row in the plot represents a pair of blended DIBs and is divided to Line1 and Line2. One can see in the second row the DIBs $5780.6$\,\AA\, and $5779.5$\,\AA\,, the pair is represented by the lowest blue point on the left in figures \ref{f:blend_sim_vals} and \ref{f:blend_sim_errs} (since their measured correlations are -0.98 and -0.96 respectively), therefore the scatter in the initial correlations is very low and their correlation uncertainty is also low (see table \ref{t:corr_red_strong}). However, the last row represents the DIBs $6845.3$\,\AA\, and $6846.6$\,\AA, which have measured correlations that are lower than 0.5 and are represented by a blue point with a red cross. One can see that the initial correlation in the bin that fits this pair is very scattered, their initial correlation uncertainty is therefore high (see table \ref{t:corr_red_strong}).

\section{Pairwise correlation simulations}\label{s:flux_flux_sims}

The algorithm described in section \ref{s:algo} consists of several phases, each phase can be executed in more than one way. We therefore use simulated absorption lines to test different approaches towards each phase. Furthermore, we use the simulations to quantify the quality of every resulting group and find parameters that affect the goodness of the clustering using a scoring system for the resulting groups. We infer subsequently thresholds to remove false-positive groups.

We simulate a number of `carriers': each carrier consists of 2-6 absorption lines (Gaussian functions), it has a correlation with color excess as well as a correlation with other carriers. The correlations, which are taken from a uniform distribution in the range [-1, 1], are expressed through the amplitude of the absorption lines similarly to the simulations in section \ref{s:sin_syn}. We construct the base amplitude of the carrier as follows:

\begin{subequations}\label{eq:9.5}
\begin{align}
\hat{A} =& \rho_{d}\cdot\mathrm{E}(B-V) + \rho_{h}\cdot H + \Delta(\rho_{d}, \rho_{h})\\
H =& \mathcal{N}(0,\sigma_{\mathrm{E}(B-V)}) \nonumber \\
\Delta =& \sqrt{1+\rho_{d}^{2}}\cdot\mathcal{N}_{d}(0,\sigma_{\mathrm{E}(B-V)}) \nonumber \\
&+ \sqrt{1+\rho_{h}^{2}}\cdot\mathcal{N}_{h}(0,\sigma_{\mathrm{E}(B-V)}) \nonumber
\end{align}
\end{subequations}

Where $\rho_{d}$ is the correlation with color excess and $\mathcal{N}(0,\sigma_{\mathrm{E}(B-V)})$ is a Normal distribution scaled to the color excess distribution. The correlation with other carriers is expressed through a correlation, $\rho_{h}$, with a hidden variable, H, which is also drawn from a Normal distribution that is scaled to the color excess. $\Delta(\rho_{d}, \rho_{h})$ represents the noise.
We assume that lines that belong to the same carrier correlate perfectly. We therefore scale their amplitude, $A_{i}$, according to the base amplitude of their carrier:

\begin{equation}\label{eq:12}
A_{i} = \hat{A}\cdot a_{i} + b_{i}
\end{equation}

Where the index $i$ can be 0 -- 6 and represents the index of the absorption line, $a_{i}$ and $b_{i}$ are the slope and the intersect with the axis and they are sampled from a log-uniform distribution in the range of [-4, 0] for every absorption line.
The standard deviation of every Gaussian is chosen from a uniform distribution in the range [1.06\,\AA, 2\,\AA] and the central wavelength is chosen from a uniform distribution in the range [3817\,\AA, 9206\,\AA] as described in section \ref{s:sin_syn}.

We divide the simulation to two separate simulations, the first simulation includes the carriers with 2 and 3 absorption lines and the second simulation includes the carriers with 4 -- 6 absorption lines. In the first simulation we inject 10 carriers with 2 absorption lines and 10 carriers with 3 absorption lines, which result in 50 injected lines per spectrum, while in the second simulation we inject 5 carriers with 4,5 and 6 absorption lines, 75 lines per spectrum. 
We inject the lines to the original 1.5M spectra and repeat the process of normalization, interpolation and stacking as discussed in section \ref{s:data} and obtain 250 coordinate-based binned spectra. We then execute the algorithm we describe in section \ref{s:algo}, we use the central wavelengths of the synthetic lines to obtain the reduced normalized correlation matrix and we perform Hierarchical Clustering on the matrix. We extract the clusters which form under a threshold of $3\sigma$. 

\begin{figure*}
\includegraphics[width=0.8\textwidth]{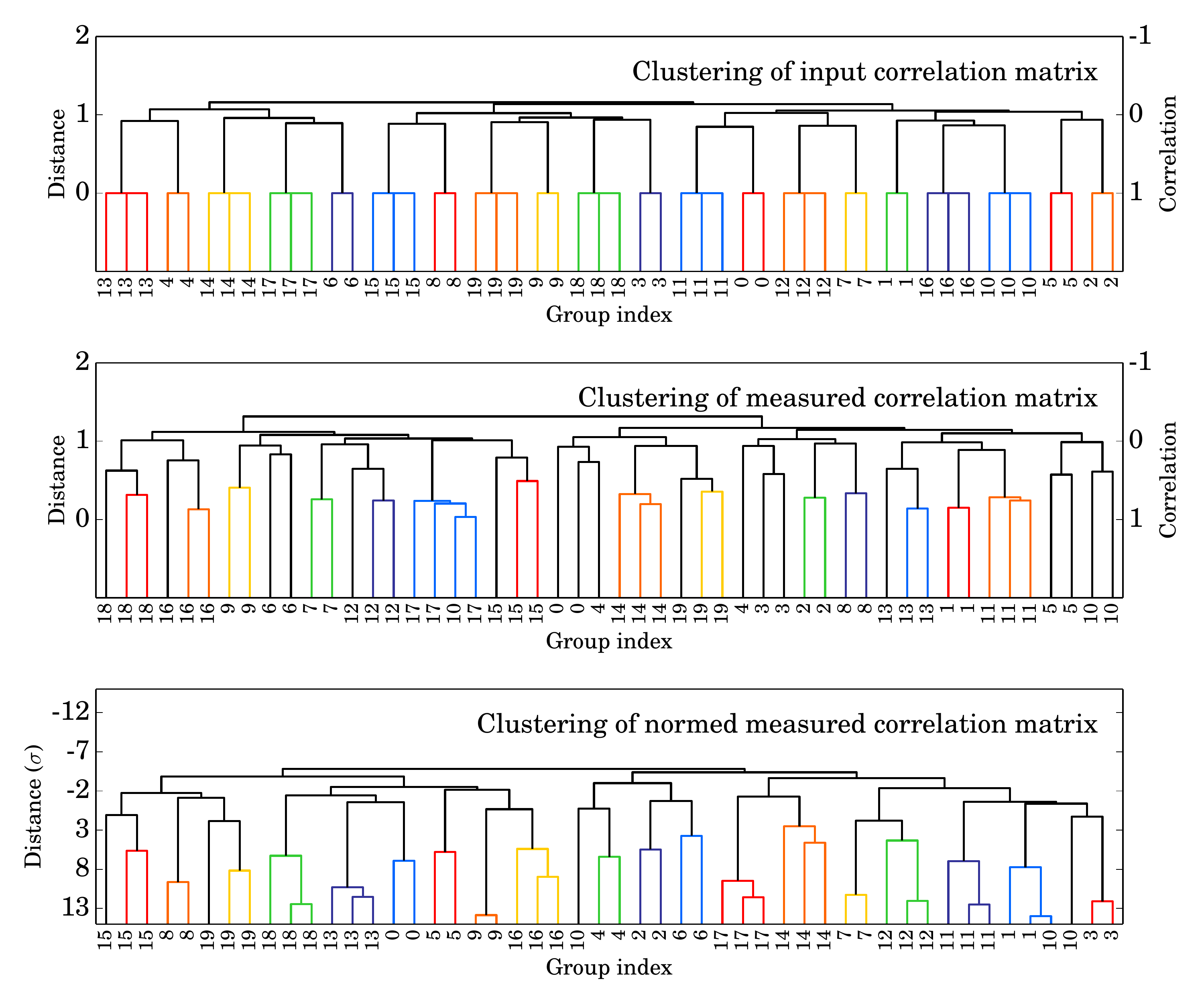}
\caption{In the top panel we show an example of input carriers we simulate, each having 2--3 lines. In the middle panel we show the clustering result using the measured correlation matrix. The bottom panel shows that using the normalized matrix improves the clustering significantly and we recover most groups, with little confusion.}\label{f:clustering_example}
\end{figure*}

In order to quantitatively compare between the resulting groups and the initial groups we define a score for every group: (1) score = `0': the final group contains the same lines as the initial group i.e., the group was recovered perfectly, (2) score = `1': the final group contains some of the lines from the initial group and does not contain additional lines i.e., the group was partially recovered and (3) score = `2': the final group contains lines that are not a part of the initial group i.e., the group was not recovered. The third score represents the false positives sample. We present an example of the clustering result for synthetic lines in figure \ref{f:clustering_example}. The top panel shows the clustering of the input lines, the middle panel shows the clustering of the measured correlations ($\rho$) and the bottom shows the clustering of the normalized correlations ($\rho_{norm}$), the labels on the x-axis represent the initial group index of every line. The newly formed group (17, 17, 10, 17) in the middle panel is an example of lines that were clustered incorrectly because of wavelength proximity since the pair of lines from the $17th$ and $10th$ group are located at 5297\,\AA\, and 5299\,\AA\, respectively. Furthermore, we find that 74\% of the false positives in the measured correlation clustering are caused by a proximity of otherwise non-correlating lines, this validates the use of the normalized correlation matrix as the basis for the clustering. One can find the three different group scores we define when comparing the top panel to the bottom panel: the group (13, 13, 13) exists in both of the clusterings and therefore it was recovered perfectly and its score is `0', the input group (19, 19, 19) was recovered as (19, 19) and therefore it was partially recovered and its score is `1'. The group (1, 1, 10) in the normalized clustering is a false positive since its lines do not belong to the same input group, its score is therefore `2'.

\begin{figure}
\includegraphics[width=3.25in]{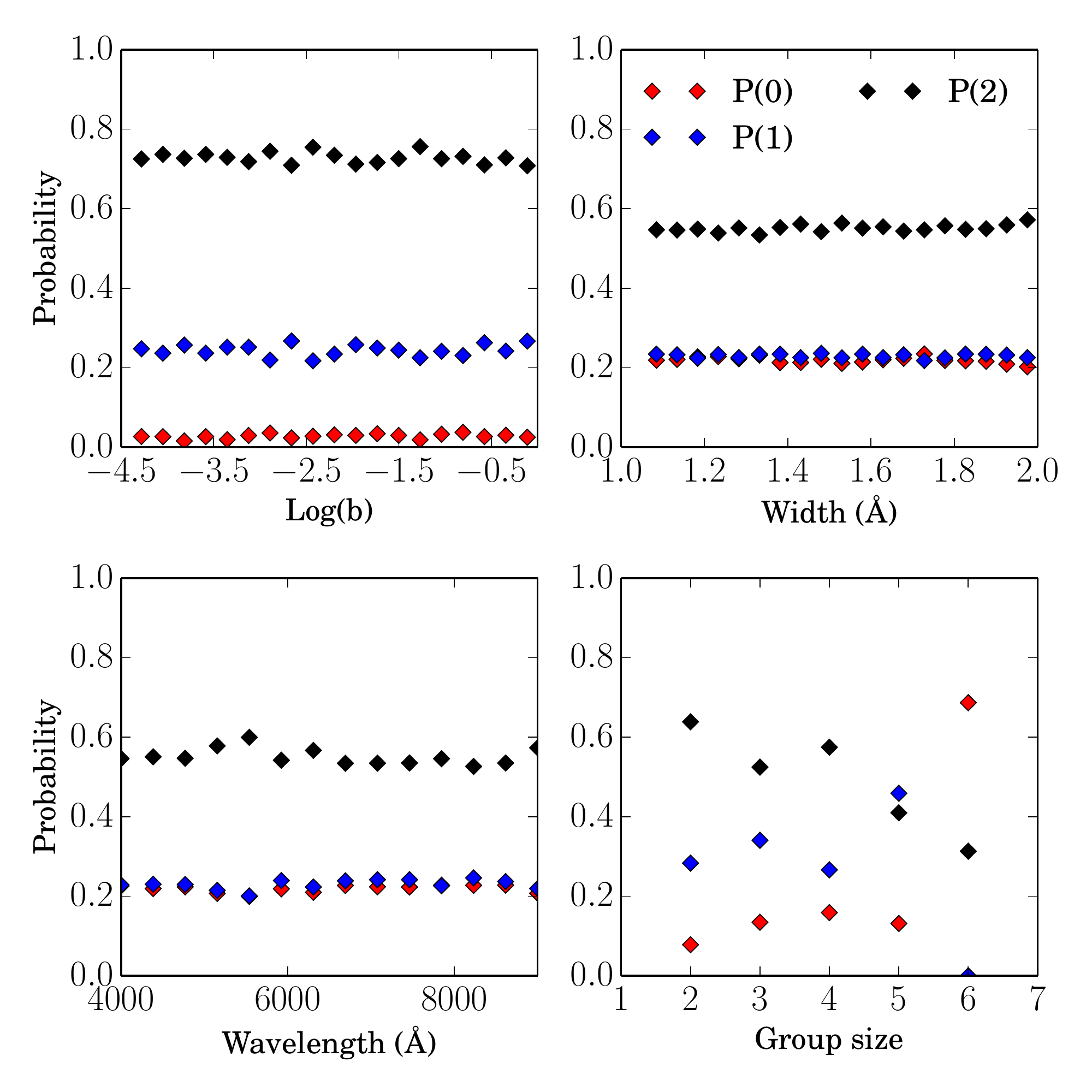}
\caption{The probability of a synthetic absorption line to be clustered in a perfectly recovered group (red), partially recovered group (blue) and incorrectly recovered group (black) as a function of the lines initial parameters: its Gaussians width and central wavelength, it intersection value with the axes and the size of its group.}\label{f:flux-flux-sims-pars}
\end{figure}

We repeat both of the separate simulations 1000 times and result with 125,000 absorption lines, times 1.5M spectra, in total. We use the lines that were clustered to groups and the groups scores in order to divide the lines and their parameters to 3 groups according to their groups score. Figure \ref{f:flux-flux-sims-pars} presents the probability of an absorption line to be clustered in a `0', `1' and `2' groups (red, blue and black respectively) as a function of the Gaussian width, central wavelength, $b$ parameter (intersection with the axis) and the size of their group. We find that the probability of a line to be clustered in a perfect, partial or false group does not depend on the Gaussians width or central wavelength. Similarly to section \ref{s:sin_syn} we find that the probability does not depend on the $b$ parameter either. The resulting group size does affect the probability, though we show in section \ref{s:dibs_groups} that the DIBs always cluster into small groups. The strongest effect we find is of the slope of the relation, and on the correlation of the line with color excess. As the slope of the relation steepens the change in the flux per bin is higher than the average scatter. Since we compute the partial flux-flux correlation we essentially remove the signal that is caused by the dust column density and remain with the signal that is perpendicular to the $\mathrm{E}(B-V)$ vector. As the correlation with reddening increases the partial correlation removes a bigger part of the signal, and one remains with a lower signal comparable to the noise in the data.

\begin{figure}
\includegraphics[width=3.0in]{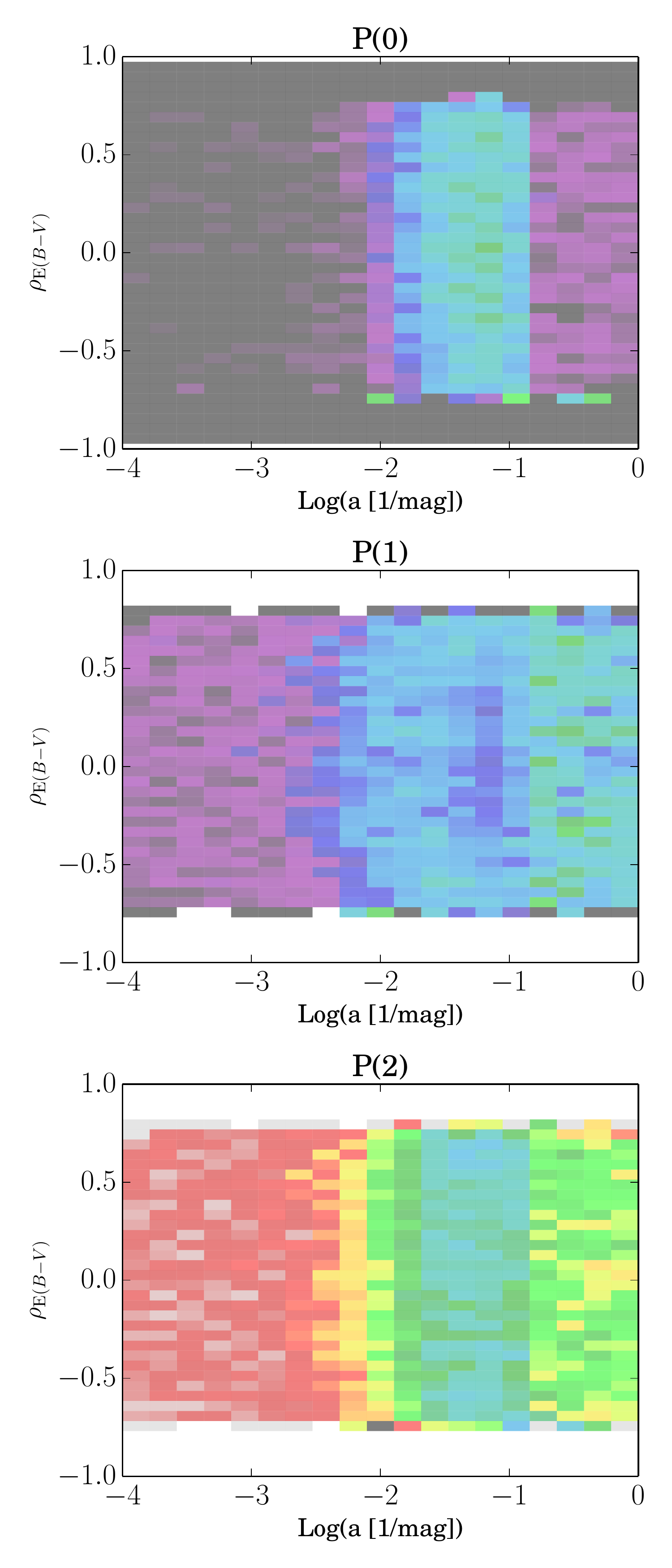}
\includegraphics[width=0.5\textwidth]{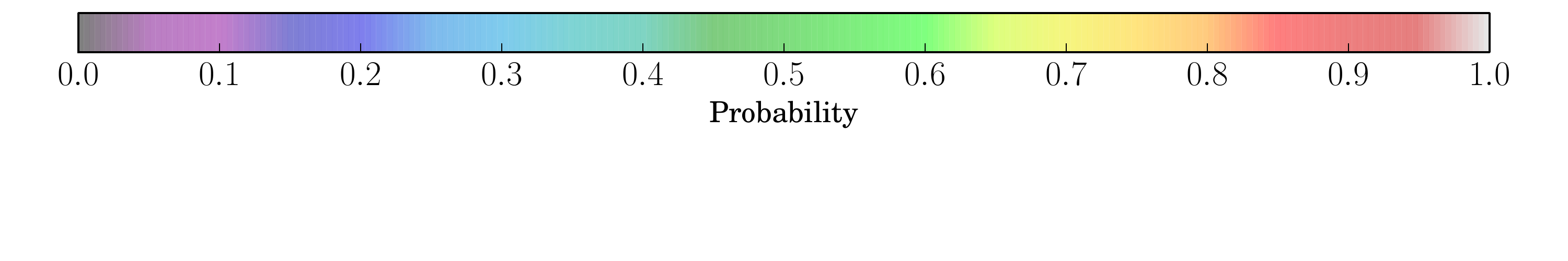}
\caption{The probability of a synthetic absorption line to be clustered in a perfectly recovered group (top panel), partially recovered group (middle panel) and incorrectly recovered group (bottom panel) as a function the slope of its flux relation and its correlation with reddening.}\label{f:prob_per_a_corr}
\end{figure}

We present in figure \ref{f:prob_per_a_corr} the probability of a line to be clustered in a `0', `1' and `2' group as a function of its slope and its correlation with $\mathrm{E}(B-V)$. The top panel in the figure presents the probability of a line to be clustered in a perfectly recovered group, the middle panel in the figure presents the probability of a line to be clustered in a partially recovered group and the bottom panel is the probability to be clustered in a false positive group. When the slope is smaller the probability of the line to be clustered in a false positive group increases, up until $a = 10^{-2.5} \frac{1}{mag}$ where the probability is almost 1 regardless of the correlation with $\mathrm{E}(B-V)$. It is also apparent that in most of the bins the probability of a line to be clustered in a partially recovered group is higher than the probability to be clustered in a perfectly recovered group.
We use the 2D binning to obtain the probabilities of the measured absorption lines (DIBs, atoms and molecules) to be clustered in `0', `1' and `2' groups in sections \ref{s:mol_groups} and \ref{s:dibs_groups}.

Since we wish to concentrate on dividing out (by probability analysis) the DIB, atom and molecule groups that were falsely clustered, we define the probability to be clustered in a false group to be our false positives (FP) and the probabilities to be clustered in `0' and `1' groups to be the true positives (TP=1-FP).
We measure the FP and TP rates as a function of the clustering method (complete, single, weighted and average) and find no dependance of the rates on the different methods. We also measure the rates for different clustering thresholds and find that a threshold of $3\sigma$ maximizes the TP/FP rate. We therefore use the complete clustering method with a $3\sigma$ threshold for the grouping.

\begin{figure}
\includegraphics[width=3.25in]{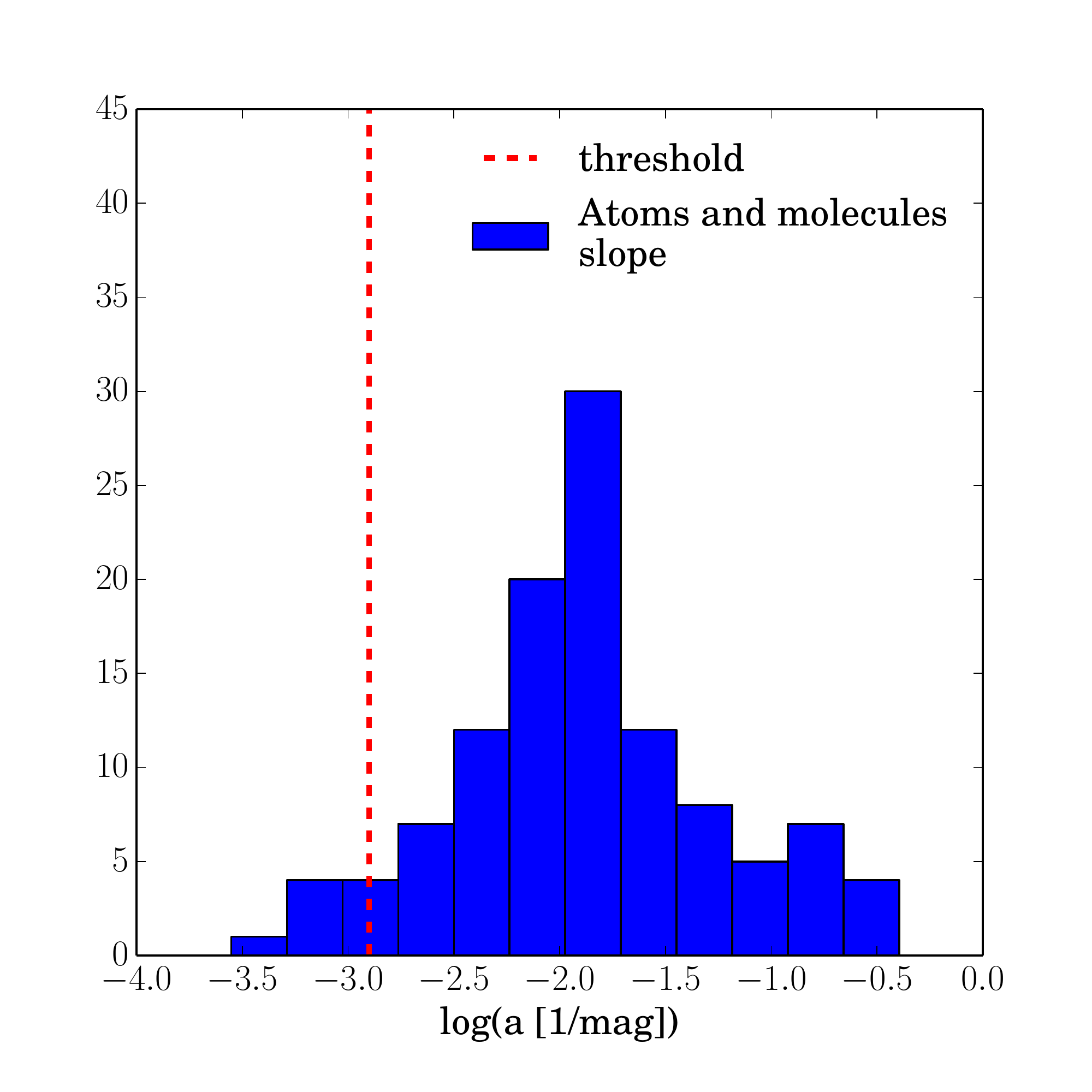}
\caption{Histogram of the slopes of atoms and molecules from our catalog (blue) and the detection threshold for the slope (red).}\label{f:atoms_slopes_hist}
\end{figure}

\begin{figure}
\includegraphics[width=3.25in]{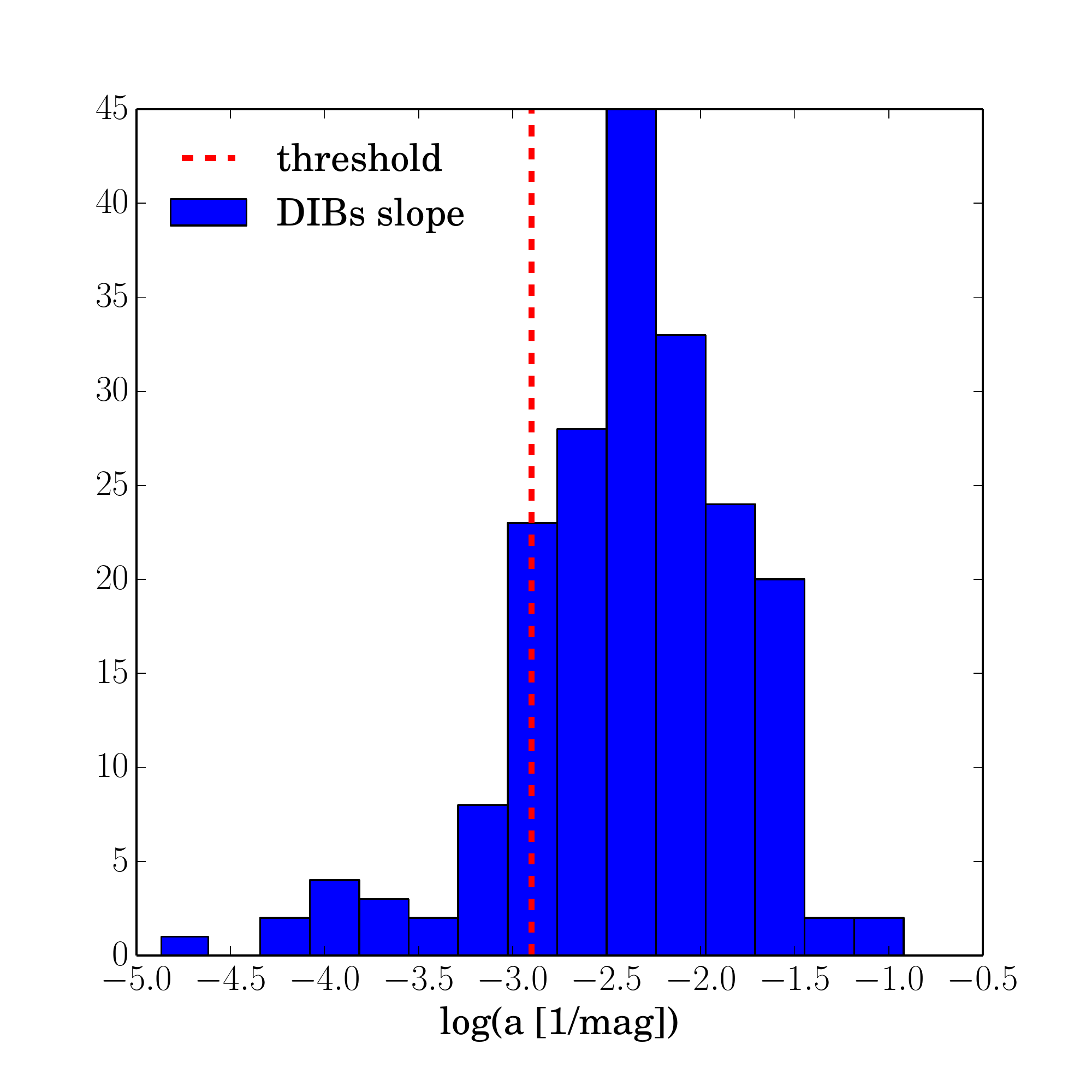}
\caption{The log-uniform distribution of DIBs slopes (blue) and the slope threshold for correlation recovery (red). DIBs with slopes smaller than -2.9 are discarded from the analysis.}\label{f:dibs_slopes}
\end{figure}

In order to assign the probability of a resulting group to be TP or FP we compute the correlation of flux with $\mathrm{E}(B-V)$ and the slope of the relation for the absorption lines. We present the slope histogram for the atomic and molecular lines in figure \ref{f:atoms_slopes_hist} and the slope histogram for the DIBs in figure \ref{f:dibs_slopes}, the threshold that we present in the figures (red) is the threshold we deduce via the simulations in section \ref{s:sin_syn} which divides out the lines we can measure a correlation with $\mathrm{E}(B-V)$ for. Clearly, most of the relations with $\mathrm{E}(B-V)$ can be measured. However, the group recovery depends also on the correlation with $\mathrm{E}(B-V)$ and we can not cluster lines that correlate strongly with dust. We then deduce the probability of every line in the clustered lines by crossmatching its slope and correlation with $\mathrm{E}(B-V)$ using figure \ref{f:prob_per_a_corr}. We find that, generally, lines that form a group share similar slopes and correlations with $\mathrm{E}(B-V)$ which is expected for lines that correlate well with each other. We therefore compute the median slope and correlation with $\mathrm{E}(B-V)$ value for every group and deduce the probability of the group to be TP or FP using figure \ref{f:prob_per_a_corr}.

\clearpage

\clearpage 
\end{document}